\documentclass[aps,prl,reprint,showpacs,superscriptaddress,longbibliography]{revtex4-2}
\usepackage{mathtools,amssymb,graphicx,units}

\usepackage[plainpages=false,pdfpagelabels,colorlinks=true,linkcolor=red,urlcolor=blue,citecolor=blue,pdftitle={Title},pdfauthor={},pdfdisplaydoctitle=true,pdfduplex=DuplexFlipLongEdge]{hyperref}
\usepackage{verbatim}
\usepackage[english]{babel}
\usepackage{color}
\usepackage{placeins}
\usepackage{tikz}
\usetikzlibrary{arrows.meta,positioning}

\usepackage[normalem]{ulem}

\renewcommand{\thesection}{\arabic{section}}
\renewcommand{\thesubsection}{\thesection.\arabic{subsection}}

\begin{document}
\include{colordefs}

\makeatletter
\def\@seccntformat#1{\csname the#1\endcsname\quad}
\makeatother

\title{Flowing to Normality and the Fate of the Single Ring Theorem}
\author{Joshua Feinberg}
\affiliation{Department of Physics, University of Haifa, Haifa 31905, Israel}
\author{Roman Riser}
\affiliation{Department of Mathematics, Tulane University, New Orleans, LA 70118, USA}
\affiliation{School of Mathematical Sciences, Holon Institute of Technology, Holon 5810201, Israel}
\author{Richard Scalettar}
\affiliation{Department of Physics and Astronomy, University of California, Davis, CA 95616, USA}
\author{A. Zee}
\affiliation{Kavli Institute for Theoretical Physics, University of California, Santa Barbara, California 93106, USA}

\date{\today}

\begin{abstract}
{Random non-hermitian matrix ensembles with double-sided rotation invariance obey, in the limit of large matrix size, the Single Ring Theorem, which states that the support of the mean eigenvalue distribution in the complex plane is either a disk or an annulus. In contrast, rotational-invariant random normal matrix ensembles can have mean eigenvalue densities supported over any number of concentric annuli in the complex plane. In this paper we introduce and investigate, both analytically and numerically, a non-hermitian matrix model which flows from a generic matrix distribution obeying the Single Ring Theorem to a distribution of normal matrices by tuning a parameter which penalizes non-normality. 

We observe numerically breakdown of the Single Ring Theorem as the model flows towards normality, and determine the critical value of the parameter at which the transition occurs. We also study in detail the behavior of the singular values of these matrices under the flow. These singular values form a Fermi gas confined to the positive half-line. In particular, we find that at small values of the flow parameter, the interparticle spacings in the gas exhibit Wigner-Dyson repulsion, whereas for asymptotically large values of the flow parameter, at the normal matrix endpoint of the flow, the spacing statistics is Poissonian. The flow interpolates continuously between these two types of statistics. However, this change in statistics is not related directly to breaking of the Single Ring Theorem, which occurs very early-on along the flow, in the regime of Wigner-Dyson statistics. Finally, we introduce a certain ensemble of random permutations associated with the gas, and make a conjecture on how to use it in order to reconstruct approximately the average density of complex eigenvalues from that of the singular values in the large-$N$ limit.}
\end{abstract}

\maketitle

\section{Introduction}\label{Section:Introduction}

Consider probability ensembles of $N\times N$ non-hermitian random matrices $\phi$ of the form 
\begin{equation}\label{prob}
    \tilde P(\phi,\phi^\dagger)  = \frac{1}{\zeta_N}e^{-N{\rm Tr}V(\phi^\dagger\phi)}\,,
\end{equation}
where $V(\phi^\dagger\phi)$ is a generic polynomial with real coefficients and
\begin{equation}\label{partition0}
    \zeta_N = \int d\phi d\phi^\dagger e^{-N{\rm Tr}V(\phi^\dagger\phi)}
\end{equation}
is the partition function. This ensemble is invariant under two-sided unitary transformations $\phi\rightarrow U\phi V$, with {\em independent} unitaries $U,V\in U(N)$. Consequently, the support of the average density of eigenvalues of matrices drawn from \eqref{prob} in the complex plane must be rotational invariant. 

The {\em Single Ring Theorem} (SRT)\cite{feinberg1997non} asserts that in the large-$N$ limit, this rotational invariant support must be either a disk or an annulus, {\em independently} of the number of minima of $V(\phi^\dagger\phi)$. 

The SRT may appear counter-intuitive at first sight. Indeed, consider a potential $V(\phi^\dagger\phi)$ with several wells or minima. For deep enough wells, one might expect the eigenvalues of $\phi^\dagger\phi$ to ``fall into the wells", consequently causing the eigenvalue distribution of $\phi$ to be bounded by a set of concentric circles of radii $0\leq r_1<r_2\ldots <r_{n_{\rm max}}$, separating annular regions in the complex plane on which the density of eigenvalues of $\phi$ is positive, from annular voids where it vanishes. Moreover, it is natural to assume a priori that the maximal number of such circular boundaries should grow with the degree of the polynomial $V$, because $V$ may have then many deep minima. Remarkably, however, according to the SRT, the number of these boundaries is two at the most. 

A compelling argument reconciling this fact with the a priori expectation mentioned above, based on the singular value decomposition (SVD) of $\phi$ was given in \cite{feinberg1997non, feinberg2001single, feinberg2006Stellenbosch}. It asserts that while the eigenvalues of the hermitian matrix $\phi^\dagger\phi$ may split into several disjoint segments along the positive real axis, this does not necessarily constrain the complex eigenvalues of $\phi$ to condense into corresponding annuli. Moreover, that argument also suggests under which conditions the SRT may be violated. 

In this paper we verify that argument and its prediction about breaking the SRT in a particular matrix model. To this end, and also to set some notation, let us first pause to recall the SVD of $\phi$: The hermitian positive matrix $\phi^\dagger\phi$ can always be diagonalized $\phi^\dagger\phi=V^\dagger \Lambda^2 V$ by a unitary matrix $V$, with $ \Lambda^2 = {\rm diag} (\lambda_1^2, \lambda_2^2,\ldots, \lambda_N^2)$ where the $\lambda_i$ are all real. Similarly, the hermitian positive matrix $\phi\phi^\dagger$, which is obviously isospectral to $\phi^\dagger\phi$, can be diagonalized $\phi\phi^\dagger = U\Lambda^2 U^\dagger$ by another unitary matrix $U$. From this follows the singular value decomposition  
\begin{eqnarray}\label{SVD}
    \phi &=& U\Lambda V\,,\quad U, V \in U(N)\nonumber\\
    \Lambda &=& {\rm diag} (\lambda_1, \lambda_2,\ldots, \lambda_N)\,, \lambda_i \geq 0\,.
\end{eqnarray}
of $\phi$. $U$ and $V$ are not unique, and are determined up to ``gauge transformation"
\begin{equation}\label{gauge}
 V\sim \Theta V, U\sim U\Theta^\dagger\,,  \Theta\in U(1)^N 
\end{equation}
Moreover, the $2^N$-fold sign ambiguity in $\Lambda$ can be fixed by absorbing the signs of individual $\lambda_i$'s into one of the unitary matrices $U$ or $V$. Thus, $\Lambda$ in \eqref{SVD} is chosen to be positive, and these positive $\lambda_i$ are known as the singular values of $\phi$ (and of $\phi^\dagger$). Let us check that parameter counting in \eqref{SVD} works: $2N^2$ (real parameters in $\phi$ )  = $N^2 + N^2$ (real independent parameters in the unitary matrices $U,V$) +$N$ (real parameters in $\Lambda$) - $N$ (real parameters in $\Theta$ that are ``gauged away"). 

Let us return to our heuristic explanation of the SRT. According to \eqref{SVD} we can always write 
\begin{equation}\label{orbit}
\phi = U(\Lambda W)U^\dagger
\end{equation}
with $W=VU$. Thus, for a fixed $W$, and as $U$ wanders freely throughout $U(N)$, we can think of 
$\phi$ as a family of unitarily equivalent matrices which are isospectral to the fiducial matrix \begin{equation}\label{fiducial}
\phi_f = \Lambda W. 
\end{equation}
Thus, it is enough to study the statistics of complex eigenvalues of $\phi_f$ for all possible $\Lambda$'s and $W$'s. Let us further assume that $V(\phi^\dagger\phi)$ has several deep minima, causing the eigenvalues of $\Lambda^2$ to condense in multiple disjoint segments around those minima. Evidently, as $W$ ranges over the unitary group $U(N)$ (which is what we expect to happen in the generic non-hermitian case), the eigenvalues of $\Lambda W$ could be smeared, in the sense that they would not span narrow annuli around the circles in the complex plane determined by the 
minima of $V(\phi^\dagger\phi).$ This idea was further pursued in detail in \cite{fyodorov2008}. 

The SRT was originally proved in \cite{feinberg1997non} by summing planar Feynman diagrams. An alternative derivation of the SRT was given later in \cite{fyodorov2007}. For a mathematically rigorous treatment see \cite{guionnet2011single}. For further developments see \cite{Nowak, Nowak1}.

The SRT was {\em demonstrated} analytically and numerically for the first time in \cite{feinberg2001single} for the concrete potential 
\begin{equation}\label{cubic1}
V(\phi^\dagger\phi)  = m^2\phi^\dagger\phi +\frac{\alpha}{2}(\phi^\dagger\phi)^2 +\frac{u}{3}(\phi^\dagger\phi)^3\end{equation} 
with $\alpha <0$. This cubic potential is the simplest non-trivial case for which the SRT applies, because for $\alpha$ negative enough, it has two competing minima in which the singular values of $\phi$ can condense in two separate bands.

The argument in favor of the SRT presented above clearly breaks down when $W$ fails to range over $U(N)$ entirely, which occurs when the matrices $U$ and $V$ are correlated\cite{feinberg2001single,feinberg2006Stellenbosch}. For example, $\phi_f$ might be such that $W$ is block-diagonal, with the upper diagonal block being a $K\times K$ unitary diagonal matrix ${\rm diag}(e^{i\omega_1},\ldots,e^{i\omega_K})$ (and with $K$ a finite fraction of $N$). In the extreme case $K=N$, in which $W$ is completely diagonal, $W=e^{i\omega} = {\rm diag}(e^{i\omega_1},\ldots,e^{i\omega_N})$, we see that $\phi = U \Lambda e^{i\omega}U^\dagger$
is a {\em normal} matrix (that is, $[\phi,\phi^\dagger]=0$), with eigenvalues ${\rm diag}(\lambda_1 e^{i\omega_1},\ldots,\lambda_N e^{i\omega_N})$, whose moduli are determined by the $\lambda_i$. Thus, normal matrices, or partially normal matrices (i.e., the case $K<N$), evade the SRT: If the first $K$ eigenvalues $\lambda_1^2,\ldots, \lambda_K^2$ of $\phi^\dagger\phi$ split into several disjoint segments along the positive real axis, the corresponding eigenvalues of $\phi$ will split into concentric annuli in the complex plane, obtained by revolving those $\lambda$-segments (see e.g.\cite{AmeurCC2025} for a recent example). 

Normal, or partially normal matrices, are of course extremely rare in the ensembles of non-hermitian matrices \eqref{prob} studied in \cite{feinberg1997non} and \cite{feinberg2001single}, and cannot affect the SRT spectral behavior of those ensembles in the large-$N$ limit. Therefore, in order to study the effect of increased matrix `normality', that is, the effect of increasing correlations between the unitaries $U$ and $V$ on the density of eigenvalues of $\phi$, we add to the potential $V(\phi^\dagger\phi)$ in \eqref{prob} a term which will bias the probability distribution in favor of normal matrices. 

The commutator $[\phi,\phi^\dagger]$ is a hermitian matrix, which vanishes when $\phi$ is a normal matrix. Thus, we shall use the {\em positive} quantity $ 0\leq   {\rm Tr}[\phi,\phi^\dagger]^2  = 2{\rm Tr}\left((\phi^\dagger\phi)^2-\phi^2\phi^{\dagger 2}\right)$
as a penalizing term for non-normal matrices $\phi$, and study in this paper the statistics of matrices drawn from the rotational invariant ensemble
\begin{equation}\label{prob1}
    P(\phi,\phi^\dagger)  = \frac{1}{\mathcal{Z}_N} e^{-N{\rm Tr}V(\phi^\dagger\phi) -Ng{\rm Tr}[\phi,\phi^\dagger]^2 }\,,
\end{equation}
with $g\geq 0$ and where 
\begin{equation}\label{partition1}
    {\mathcal{Z}_N} = \int d\phi d\phi^\dagger e^{-N{\rm Tr}V(\phi^\dagger\phi) -Ng{\rm Tr}[\phi,\phi^\dagger]^2 }
\end{equation}
is the associated partition function. Note that both terms in the exponent of \eqref{prob1} are of order $N^2$, rendering the penalty term relevant for finite values of $g\sim {\cal O}(N^0)$.

The positive parameter $g$ is a flow parameter: At $g=0$ we are back at the original model \eqref{prob} for which the SRT holds, and as $g$ increases the ensemble gradually favors matrices whose degree of normality (the block size $K$ mentioned above) grows. Indeed, the added term $Ng{\rm Tr}[\phi,\phi^\dagger]^2$ in \eqref{prob1} tells us that the eigenvalues of $[\phi,\phi^\dagger]$ are normally distributed with variance $1/(Ng)$. Thus, as $g\rightarrow\infty$ these eigenvalues get peaked sharply around zero, concentrating the entire ensemble on normal matrices, which maximally violate the SRT. We note in passing that the resulting limiting ensemble is not the canonical ensemble \eqref{zpdf} of normal matrices. In between these two endpoints of the flow, at some critical value $g_{\rm crit}$, we expect the shape of the average eigenvalue distribution of $\phi$ to change from single-ring to multi-ring appearance. 

\begin{figure}[t!]
    \centering
    \includegraphics[width=1\linewidth]{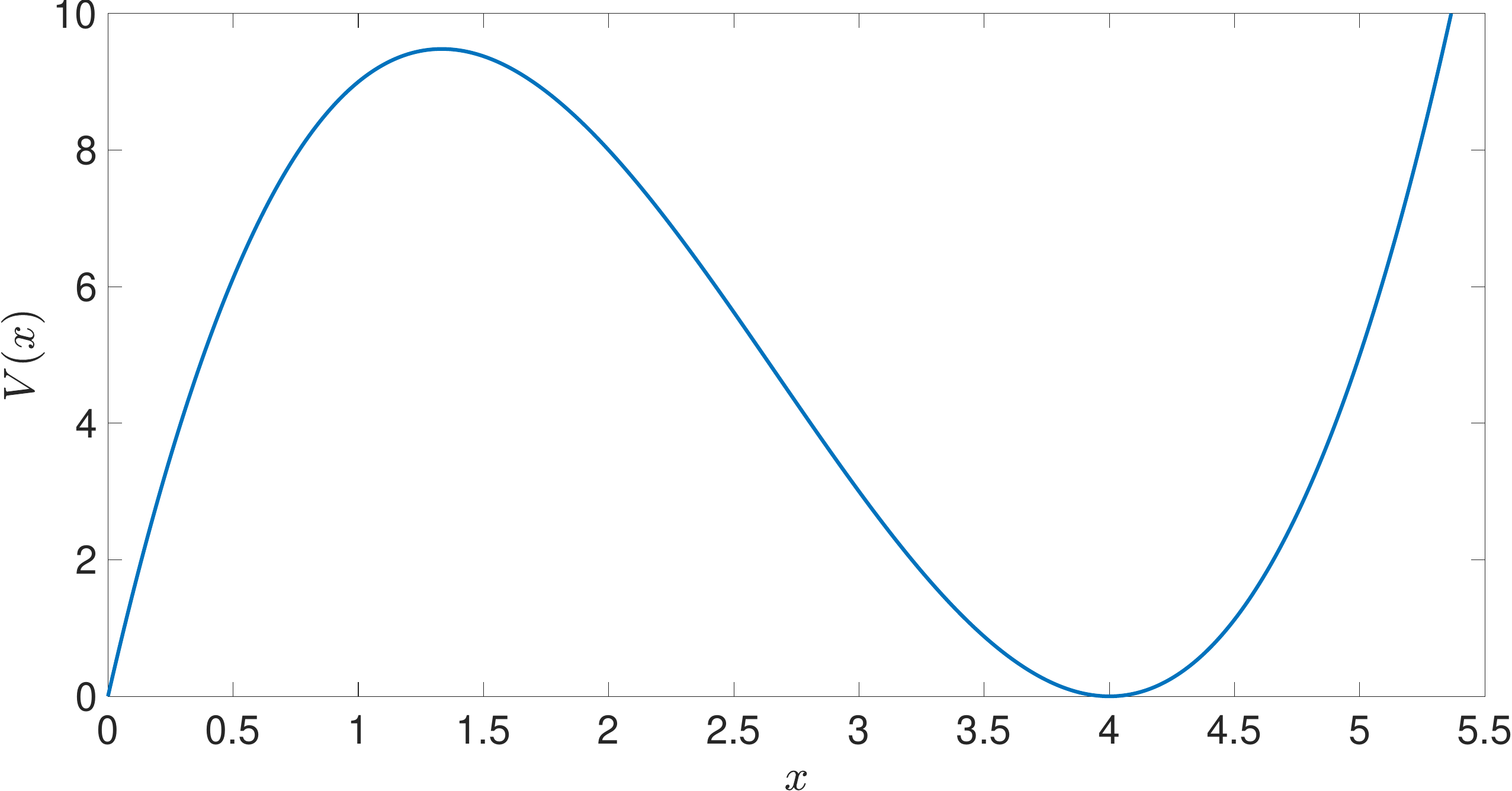}
    \caption{The cubic potential used in simulations $V(x)=m^2 x+\tfrac{\alpha}{2} x^2+ \tfrac{u}{3} x^3$ with $m^2=16$, $\alpha=-16$, $u=3$.}
    \label{fig:potential}
\end{figure}

\begin{figure}[h]
    \centering
     \includegraphics[width=3.40in,height=3.40in,angle=-90]{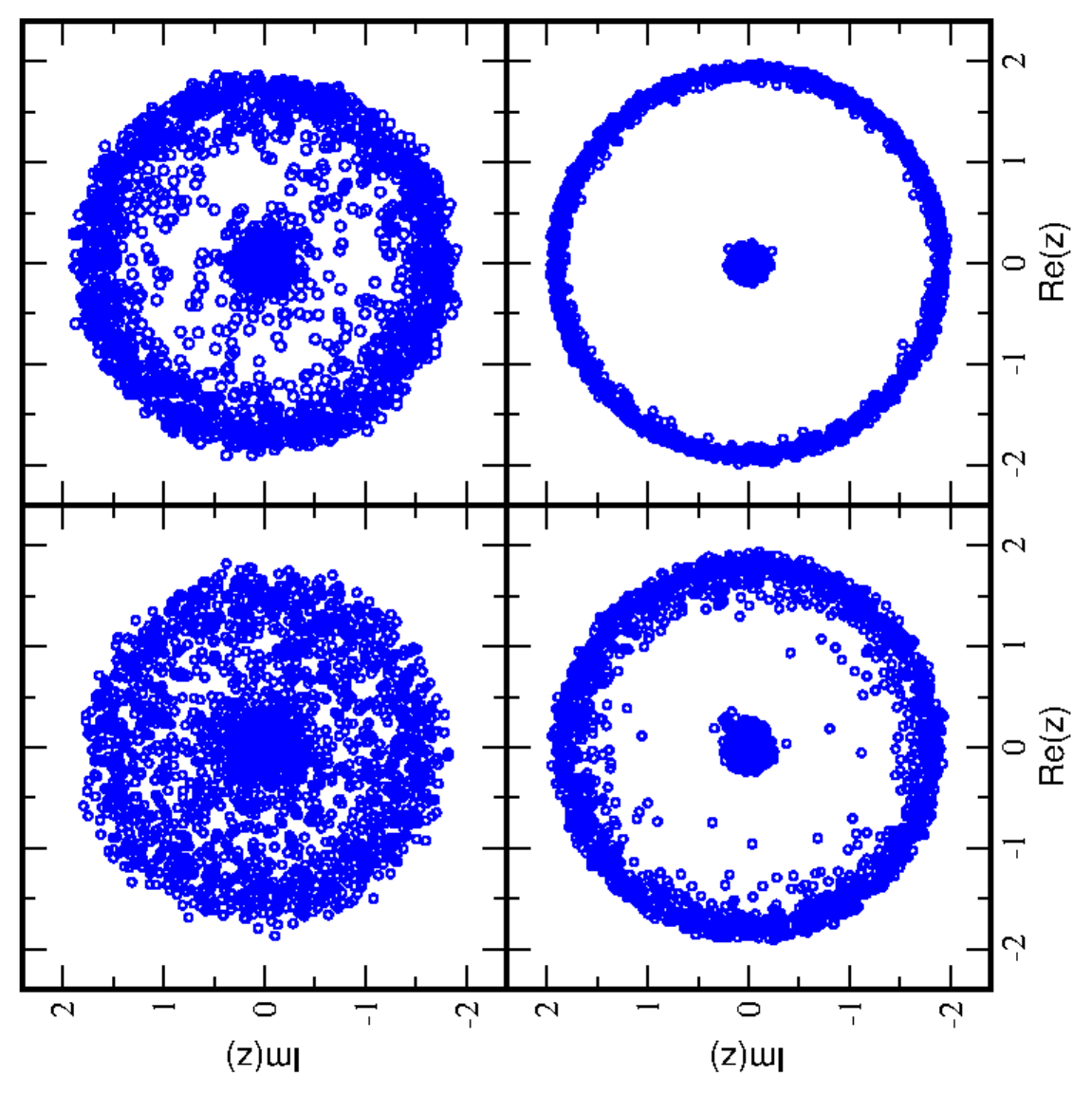}
    \caption{ Evolution of the eigenvalue distribution in the complex plane
    for the cubic potential \eqref{cubic1} with $m^2=16$, $\alpha=-16$, $u=3$. 
    Top, left: $g=0.0$. 
    Top, right: $g=0.0375$. 
    Bottom, left: $g=0.0625$. 
    Bottom, right: $g=0.125$.
    Violation of the Single Ring Theorem is evident. These data 
    suggest critical $0.0375 < g_{\rm crit} < 0.0625$.
    Matrix size $N=24$.
    }
    \label{fig:qualitative}
\end{figure}

One of our objectives in this paper is to verify this expectation numerically. This we do by performing Monte-Carlo simulations of the ensemble \eqref{prob1} with the cubic potential \eqref{cubic1}, with specific coupling constants $m^2=16, \alpha=-16$ and $u=3$. $V$ is depicted in Fig.~\ref{fig:potential} as function of an eigenvalue of $\phi^\dagger\phi$, namely the square of a singular value $x=\lambda^2$.
We turn $g$ on and carry numerical simulations of \eqref{prob1}, with this particular $V(\phi^\dagger\phi)$, for a series of increasing values of $g$. For each such value, we measure the average density of eigenvalues $z_i$ of $\phi$ in the complex plane. Some results of our measurements are shown in the scatter plots in Fig.~\ref{fig:qualitative} for four values of $g$.
Rotational invariance of the shape of the support of eigenvalues on average is evident \footnote{Sampling over infinite realizations of $\phi$ would result in perfect rotational invariant shape. Slight possible deviations from perfect rotational invariance in Fig.~\ref{fig:qualitative} are due to finite sampling.}.
At the start, for $g=0$, as in Section V of \cite{feinberg2001single}, in the large-$N$ limit, the singular values of $\phi$ condense in two separate disjoint bands, one touching the endpoint minimum at the origin, and the other around the minimum at $x=4$. (See the curve corresponding to $g=0$ in Fig.~\ref{fig:SVdensity} below.) Nevertheless, the complex eigenvalues of $\phi$ (top left panel in Fig.~\ref{fig:qualitative}) occupy a disk centered at the origin (with non-monotonic radial density profile), in accordance with the SRT. 

As $g$ increases, the density is depleted in an annulus about the origin (top right panel of Fig.~\ref{fig:qualitative}). Finally, at the largest $g$ shown (bottom right panel), there are two disjoint regions -- a disk and a ring both centered at the origin, but with vanishing density in between. The SRT clearly does not hold for this latter case. The bottom left panel corresponding to $g=0.0625$ also shows such two disjoint concentric regions, but with diffusive edges which should be interpreted as finite-$N$ corrections. It is already in the SRT-broken phase, a conclusion that is further supported by the refined numerical results presented in Section \ref{sec:results}. Thus, the critical value $g_{\rm crit}$ at which the SRT is violated for the first time clearly occurs somewhere in the range $0.0375<g_{\rm crit}<0.0625$ and is rather small, rendering the SRT rather fragile under the perturbation ${\rm Tr}[\phi,\phi^\dagger]^2$. 

In this paper we study this phenomenon numerically in detail and estimate the critical value $g_{\rm crit}$. This we do in the second part of the paper (see sections \ref{sec:methods} and \ref{sec:results}). In particular, in the end of Section \ref{sec:results} we give a better numerical estimate $g_{\rm crit}=0.055\pm0.005$, expected to be valid in the large-$N$ limit. 

Some aspects of our Monte Carlo analysis of the distribution of complex eigenvalues of $\phi$ involve also information gleaned from the numerically measured distribution of the singular values of the model. Therefore, in the first part of this paper we discuss the gas of singular values of these matrices (see sections \ref{sec:svgas} and \ref{sec:Fermi}). These singular values form a Fermi gas confined to the positive half-line. We derive their exact joint probability distribution function (JPDF), from which we compute numerically their average density for various values of $N$. As can be seen in Fig.~\ref{fig:SVspdistN2x} below, for small $g$, in a range including the critical value $g_{\rm crit}$, this gas exhibits Wigner-Dyson repulsion (in the GUE universality class) among the singular values. However, for asymptotically large $g$ repulsion is suppressed, leading (on a distance scale much larger than $1/\sqrt{Ng}$) to Poissonian interparticle spacing behavior, similarly to \cite{MNS}. Thus, the change in singular value statistics as $g$ is increased has nothing to do with the fragility of the SRT under the flow, as the SRT gets broken already at a rather small $g_{\rm crit}\approx 0.05$. 

\begin{figure}[h]
    \centering
\resizebox{\columnwidth}{!}{
\begin{tikzpicture}[>=Stealth,scale=1.2]
 \draw[thick] (0,0) -- (6.21,0);
 \draw[thick,->] (6.3,0) -- (9,0) node[right] {$g$};
 \node at (6.25,0) {$\slash \! \slash$};

 \draw (0,0.2) -- (0,-0.2) node[above=0.4cm] {$g = 0$};
 \draw (1.6,0.2) -- (1.6,-0.2) node[above=0.4cm] {$\phantom{{}_\text{crit}}g = g_{\text{crit}}$};
 \draw (8.6,0.2) -- (8.6,-0.2) node[above=0.4cm,align=center] {$N \gg 1$, $g \gg 1$ \\[0.5em]    $\sqrt{N/g} \ll 1$ };

 \node[below=0.1cm] at (0.8,0) {SRT valid};
 \node[below=0.1cm] at (2.6,0) {SRT broken};
 \node[below=0.2cm] at (1.6,0) {$\;\;\;\; \underbrace{\phantom{\text{SRT valid \ \;\; SRT broken}}}_{\substack{\phantom{text} \\ \textstyle{\text{WD statistics}}}}$};

 \node[below=0.1cm,align=center] at (8,0) {normal matrix limit \\ i.i.d. particles \\ Poisson statistics};
\end{tikzpicture}
}
    \caption{ Schematic behavior of our matrix model along the $g$-flow. The SRT holds formally only at $g=0$. By continuity, the shape of the support of eigenvalues in the complex plane is still a `single ring' (a disk) in the range $g<g_{\rm crit}$, and then splits into concentric disk and annulus for $g>g_{\rm crit}$, where the SRT is broken. Interparticle repulsion statistics starts as Wigner-Dyson at small $g$ and continuously crosses over to i.i.d.~Poisson behavior at the asymptotic normal-matrix endpoint of the flow.  
    }
    \label{fig:schematic}
\end{figure}
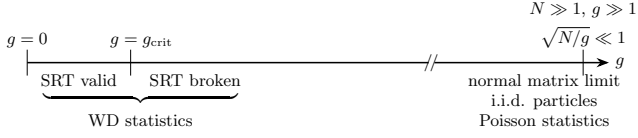

Thus, as our matrix model varies under the flow of the parameter $g$, two phenomena occur, which are not directly related. The first one is breaking of the SRT, which occurs at a finite (${\cal O}(N^0)$) small value $g_{\rm crit}$. The other one is the continual crossover of interparticle spacings from Wigner-Dyson statistics at small $g$, to non-repulsive i.i.d.~Poisson statistics at the normal matrix $g\gg N\gg 1$ endpoint of the flow. A schematic description of the behavior of our model under this $g$-flow is depicted in Fig.~\ref{fig:schematic}.

Finally, we introduce in Section \ref{sec:conjecture} a certain ensemble of random permutations associated with the distribution of of singular values, and make a speculative conjecture on how to use it to reconstruct approximately the average distribution of the complex eigenvalues of $\phi$ in the present model from the distribution of singular values in the large-$N$ limit. Several technical and formal issues are relegated to the appendices. In Appendix \ref{AppA} we discuss some mathematical aspects of normal matrices relevant for our work. In Appendix \ref{AppB} we review briefly some essential facts about the gas of free fermions on the half-line. These facts are the basis for the discussion in Section \ref{sec:Fermi}. Finally, in Appendix \ref{AppC} we derive the interparticle spacing distribution in a `gas' of $N=2$ singular values and show explicitly how the spacing distribution changes continuously along the flow from Wigner-Dyson repulsion to that of two identically independently distributed variables. It is the latter which converges asymptotically upon unfolding to Poisson statistics in the large-$N$ limit. These results are used in Section \ref{sec:svgas}.

\section{The Probability Distribution of Singular Values}\label{sec:svgas}

Let us now express the partition function \eqref{partition1} as an integral over $U,V$ and $\Lambda$. The integration measure in \eqref{partition1} can be written in `polar form' in terms of these variables as\cite{rectangles,Periwal}
\begin{equation}\label{Jacobian}
    d\phi d\phi^\dagger = d\mu(U)d\mu(V)\Delta^2(\Lambda^2) \prod_k \lambda_k d\lambda_k
\end{equation}
where $d\mu (U)$ is the Haar measure over $U(N)$ and $\Delta(\Lambda^2) = \prod_{i>j}(\lambda_i^2-\lambda_j^2)$ is the Vandermonde determinant.

From the relation 
\begin{equation}\label{positive1}
 {\rm Tr}[\phi,\phi^\dagger]^2  = 2{\rm Tr}\left(\Lambda^4-\Lambda^2 W^\dagger\Lambda^2 W\right)
\end{equation}
we see that the total potential 
\begin{eqnarray}\label{potential}
 &&{\rm Tr}\left[V(\phi^\dagger\phi) + g [\phi,\phi^\dagger]^2\right] = \nonumber\\
   &=& \sum_i \left(V(\lambda_i^2) +2g\lambda_i^4\right) - 2g{\rm Tr}(\Lambda^2 W^\dagger\Lambda^2 W)
\end{eqnarray}
in the exponent of \eqref{partition1} depends on the unitary matrices only through the combination $W=VU$. Thus, we use homogeneity of the Haar measure in \eqref{Jacobian} and change variables to $U$ and $W$. Integration over $U$ factors out and results in a factor $C_1$ of the volume of $U(N)/U(1)^N$, which we do not bother to write explicitly. (Moding out the volume of $U(1)^N$ is the result of gauge invariance \eqref{gauge}). Thus, 

\begin{eqnarray}\label{partition2}
&&\frac{\mathcal{Z}_N }{C_1}= \nonumber\\
&&\int_0^\infty \left(\prod_k \lambda_k d\lambda_k \right)\,\prod_{i>j}(\lambda_i^2-\lambda_j^2)^2\, e^{-N\sum_i\left(V(\lambda_i^2) + 2g\lambda_i^4\right)}\cdot\nonumber\\
&&\int_{U(N)} d\mu(W) e^{2Ng{\rm Tr}(\Lambda^2 W^\dagger\Lambda^2 W)}
\end{eqnarray}
The last integral in \eqref{partition2} over the unitary group can be computed from the Itzykson-Zuber formula \cite{IZ}, which yields

\begin{equation}\label{eq:IZ}
\int_{U(N)} d\mu(W) e^{2Ng{\rm Tr}(\Lambda^2 W^\dagger\Lambda^2 W)}
= \frac{C_2}{g^{N(N-1)/2}} \frac{\det_{ij}e^{2Ng\lambda_i^2\lambda_j^2}}{\Delta^2(\Lambda^2)} \,,  
\end{equation}
where $C_2$ is another normalization constant. The factor $g^{N(N-1)/2}$ in the denominator on the RHS of \eqref{eq:IZ} can be deduced from the fact that as $g\rightarrow 0$, the integral over $W$ should just produce the volume of the unitary group. Therefore, the $\lambda$-dependent part of the determinant in the numerator should cancel the denominator, and the precise exponent of $g$ then follows from the homogeneity of the Vandermonde determinant in the variables $\lambda_i^2$.

We see that the Vandermonde determinants in the denominator of \eqref{eq:IZ} knockout completely the Vandermonde determinants in \eqref{partition2}, and we end up with 
\begin{eqnarray}\label{partition3}
\mathcal{Z}_N  =
&&\frac{C}{g^{N(N-1)/2}}\int_0^\infty \left(\prod_k 2\lambda_k d\lambda_k \right)\, 
\nonumber \\
&& e^{-N\sum_i\left(V(\lambda_i^2) + 2g\lambda_i^4\right)}\,
\det_{ij}e^{2Ng\lambda_i^2\lambda_j^2}\nonumber\\
=&&\frac{C}{g^{N(N-1)/2}}\int_0^\infty \left(\prod_k 2\lambda_k d\lambda_k \right)\,
\nonumber \\
&& e^{-N\sum_i V(\lambda_i^2)}\,\det_{ij}e^{-Ng(\lambda_i^2-\lambda_j^2)^2}\nonumber\\{}
\end{eqnarray}
where we have lumped together $C_1$ into $C_1C_2 = 2^N C$, and in the equation in \eqref{partition3} we absorbed in the $i$-th row of the determinant a factor $e^{-Ng\lambda_i^4}$ and a factor $e^{-Ng\lambda_j^4 }$ in its $j$-th column. 
Thus, we can write the JPDF of the $N$ singular values very neatly as 
\begin{eqnarray}\label{jpdsv}
&&\Pi(\lambda_1,\ldots,\lambda_N)d\lambda_1\ldots d\lambda_N   = \frac{1}{\mathcal{Z}_N}\left(\prod_k 2\lambda_k d\lambda_k \right)\cdot\nonumber\\ 
&&\det_{ij}\left(e^{-\frac{1}{2}NV(\lambda_i^2)}e^{-Ng(\lambda_i^2-\lambda_j^2)^2}e^{-\frac{1}{2}NV(\lambda_j^2)}\right)\,.
\end{eqnarray}
We now change variables one last time to $x_i = \lambda_i^2$ and write \eqref{partition3} as 
\begin{eqnarray}\label{partition}
   &&\mathcal{Z}_N =\nonumber\\ 
   &&\frac{C}{g^{N(N-1)/2}} \int_0^\infty\prod_k dx_k\,e^{-N\sum_i V(x_i)}\,\det_{ij}e^{-Ng(x_i-x_j)^2}=\nonumber\\
  && \frac{C}{g^{N(N-1)/2}} \int_0^\infty\prod_k dx_k\,\cdot\nonumber\\
  &&~~~~~~~~~\det_{ij}\left(e^{-\frac{1}{2}NV(x_i)}e^{-Ng(x_i-x_j)^2}e^{-\frac{1}{2}NV(x_j)}\right)\,.
\end{eqnarray}
where in the last step we absorbed in the $i$-th row of the determinant a factor $e^{-\frac{1}{2}NV(x_i)}$ and a factor $e^{-\frac{1}{2}NV(x_j)}$ in its $j$-th column.

Thus, to summarize, the JPDF of the squares $x_i=\lambda_i^2$ of singular values is  
\begin{eqnarray}\label{jpdx}
  && P(x_1,x_2,\ldots,x_N)  =\nonumber\\
  &&\frac{1}{\mathcal{Z}_N} \det_{ij}\left(e^{-\frac{1}{2}NV(x_i)}e^{-Ng(x_i-x_j)^2}e^{-\frac{1}{2}NV(x_j)}\right)\,.
\end{eqnarray}

\begin{figure}[h]
    \centering
    \includegraphics[width=1\linewidth]{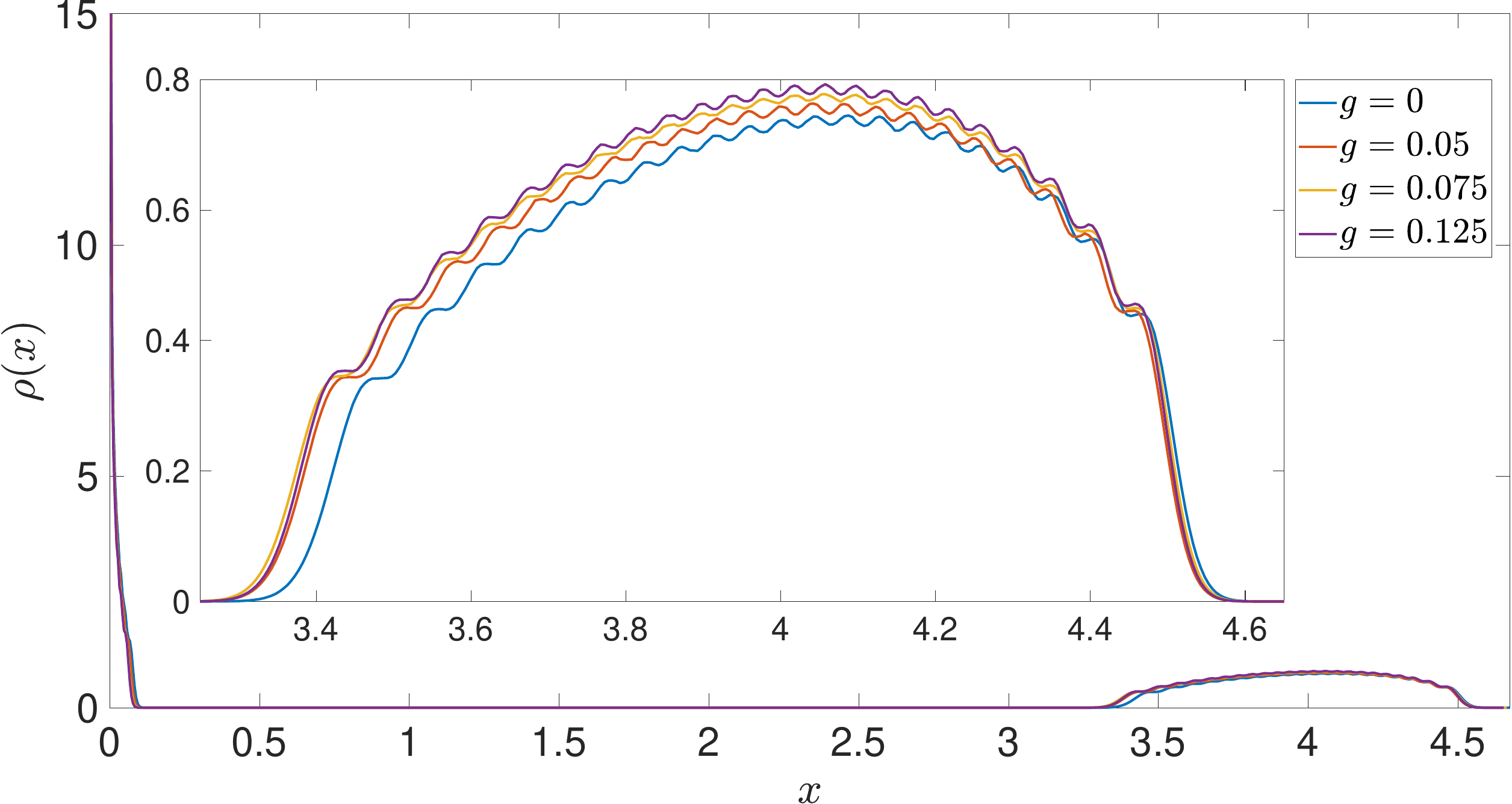}
    \caption{Density of squares singular values $x_i$ computed from \eqref{jpdx} with the cubic potential $V$ shown in Fig.~\ref{fig:potential} using Monte Carlo simulation for matrices of size $N=32$ and for $g$ value across the critical range. The singular values are clearly seen to condense in two disjoint segments. The inset magnifies the smaller density region around $x=4$. The density oscillations visible in the magnified inset are a typical finite-$N$ effect. Notice that the density barely changes as $g$ sweeps across the indicated range.}
    \label{fig:SVdensity}
\end{figure}

We have computed numerically the density of singular values 
\begin{eqnarray}\label{rhox}
  && \rho(x) = \Big\langle \frac{1}{N}\sum_{i=1}^N\delta (x-x_i)\Big\rangle =\nonumber\\
  &&\int dx_2 \ldots dx_N\, P(x,x_2,\ldots,x_N) 
\end{eqnarray}
corresponding to \eqref{jpdx} for the cubic potential $V$ plotted in Fig.~\ref{fig:potential} and for various values of $g$ ranging across the critical value $g_{\rm crit}$, for matrices of size $N=32$. The results are displayed in Fig.~\ref{fig:SVdensity}. The singular values are clearly seen to condense in two disjoint segments. Notice that the density barely changes as $g$ varies in the indicated range.

\subsection{Continuous crossover from Wigner-Dyson to Poisson Interparticle Spacing Statistics}\label{subsec:statistics-change}

In the next section (see also Appendix \ref{AppB}) we show that due to its determinantal form, the JPDF \eqref{jpdx} is in fact the diagonal matrix element (in position space) of the density matrix of a gas of $N$ non-interacting fermions at thermal equilibrium, living on the positive half-line. Pauli's exclusion principle then implies repulsion among the $x_i$: The JPDF \eqref{jpdx} vanishes whenever two such particles coincide.

As explained at the end of Appendix \ref{AppB} (for the potential-free case $V=0$), for small finite values of $Ng$, this repulsion is of the $\beta=2$ Wigner-Dyson type, as in the case $g=0$ in the presence of a confining potential $V(x)$. Indeed, in the limit $g\rightarrow 0^+$, the integral \eqref{eq:IZ} tends to a constant, and repulsion among the $x_i$'s crosses over to the conventional $\beta=2$-type Wigner-Dyson behavior (for any confining potential $V(x)$) with spacings of order $1/N$, due to the $\Delta^2(\Lambda^2)$ factor in \eqref{partition2}. In the opposite limit, as $Ng\rightarrow\infty$ towards the endpoint of normal matrices, repulsion among singular values disappears completely (over length scales larger than $1/\sqrt{Ng}$), which is also the case for the canonical ensembles of normal matrices \eqref{rotational}, as explained in Appendix \ref{AppA}. It is natural to expect Poisson level statistics to be associated with this suppressed repulsion. We argue as follows: Typical consecutive interparticle spacings are of the order $\frac{1}{N}$  or larger. Thus, for large $g\gg N$, the gaussian part of the matrix elements inside the determinant in \eqref{jpdx} becomes essentially the unit matrix $\delta_{ij}$, except for the extremely rare occasion when a finite fraction of all particles condense into much shorter separations of the order of $1/\sqrt{Ng}$. Thus, for $g\gg N$, a good approximation for \eqref{jpdx} is 
\begin{equation}\label{jpdxiid}
P(x_1,x_2,\ldots,x_N)\approx\frac{1}{\mathcal{Z}_N} \prod_{i=1}^N e^{-NV(x_i)}\,,
\end{equation} 
subjected to the constraint that pairs of particles cannot come closer than a distance $\sim 1/\sqrt{Ng}$ in space. That is, except for an excluded volume fraction of the order $N\frac{1}{\sqrt{Ng}} = \sqrt{\frac{N}{g}}\ll 1$, the gas is approximately that of identically independently distributed (i.i.d.) particles. This behavior is demonstrated by an explicit calculation in Appendix \ref{AppC} for $N=2$ particles. 

For i.i.d. particles, we know from elementary probability theory that the unfolded interparticle spacing statistics in the large-$N$ limit is {\em universally} Poissonian $P(s)=e^{-s}$ \cite{Feller1} (section 6.6), \cite{Feller2} (section 1.4). (An elementary derivation of this result for uniformly distributed particles, which is reminiscent of derivation of the survival probability in radioactive decay, is given in section 1.4 in \cite{Mehta}. For a more recent pedagogical derivation of this result for i.i.d. random variables with arbitrary probability distribution function, as in the case of \eqref{jpdxiid},
see section 2.3 in \cite{Vivo}. See also section 3.1 of \cite{RK-2021}.) Thus, an important feature of \eqref{jpdx} is that it interpolates continuously between conventional $\beta=2$-type Wigner-Dyson statistics of the $x_i$'s at $g=0$ and non-repulsive Poisson statistics at $Ng\rightarrow\infty$.

This change in statistics between the two extremes of the $g$-flow is a feature of the short interparticle distance behavior of the gas, and therefore cannot depend on the details of the potential $V(x)$.

This universal change in statistics can be demonstrated analytically in the case of the quadratic potential  $V(x)=x^2$ (that is, for the matrix potential ${\rm Tr}(\phi^\dagger\phi)^2$). In this case, our model \eqref{jpdx} coincides essentially with the restriction to the positive half-line of the JPDF of the eigenvalues of the distorted Gaussian model of hermitian matrices studied in \cite{MNS}, who mapped it on the canonical ensemble of a gas of fermionic harmonic oscillators. In fact, the original motivation in \cite{MNS} was the construction of an analytically solvable model which interpolates continuously between Wigner-Dyson statistics of eigenvalues ($g=0$) and non-repulsive Poisson statistics ($Ng\rightarrow\infty$). Thus, based on the results of \cite{MNS}, we conclude that also in our model with $V(x)=x^2$, repulsion among singular values in \eqref{jpdx} at the $Ng\rightarrow\infty$ endpoint of the flow will be suppressed, resulting in Poisson statistics of the $x_i$.

We have verified this change in statistics numerically also for the cubic potential \eqref{cubic1} and for Ginibre's potential $V(x)=x$.

\begin{figure}[h]
    \centering
    \includegraphics[width=1\linewidth]{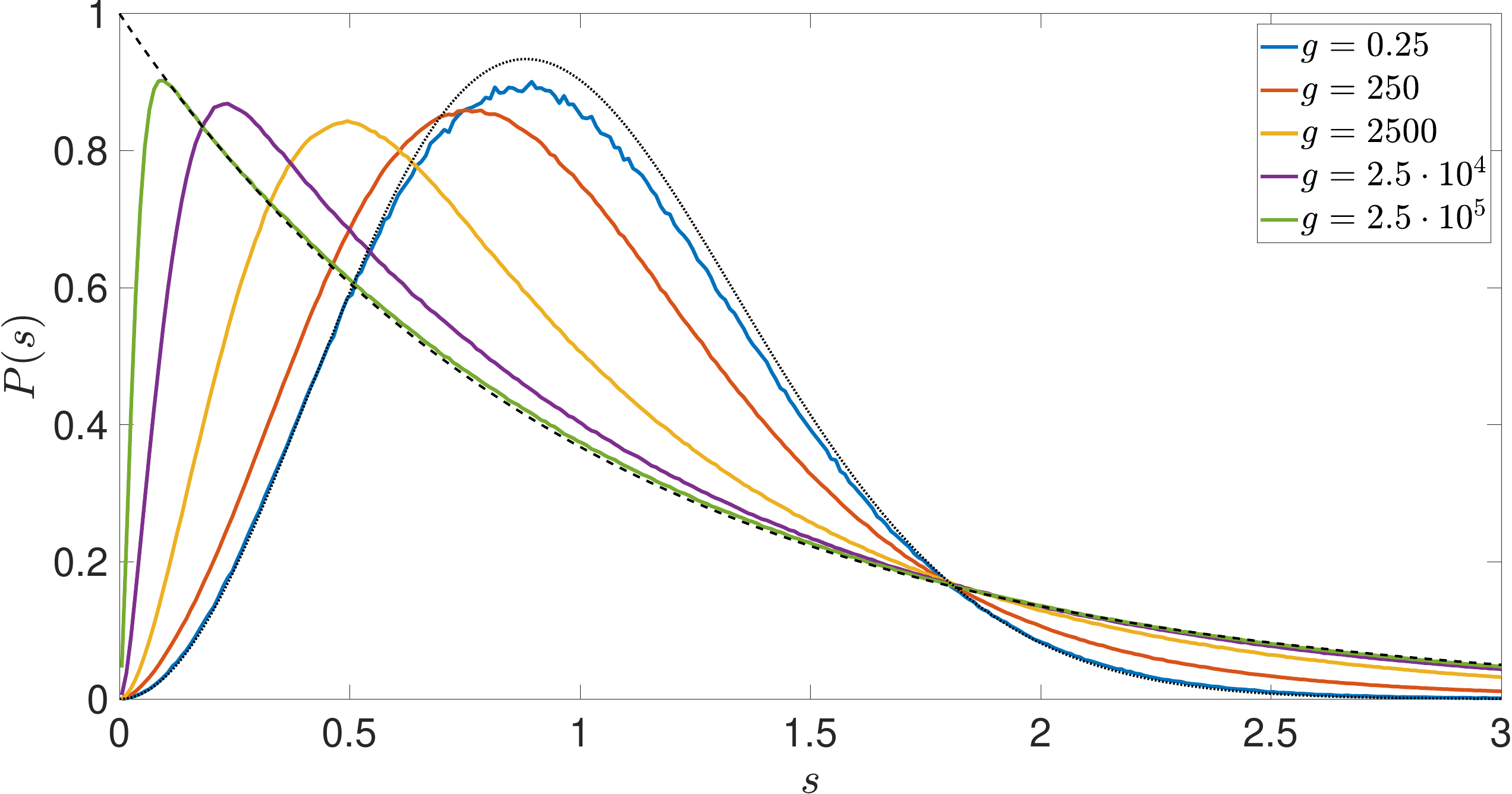}
    \caption{Unfolded spacing distribution of square singular values for various magnitude of $g$ showing transition for Wigner-Dyson ($g=0$) towards Poisson statistics. Solid lines were obtained through Monte Carlo simulation with $N=32$ singular values subjected to the critical potential (see Fig.~\ref{fig:potential}). Dashed line shows the exponential distribution $e^{-s}$ and the black dotted line represents Wigner-Dyson statistics for $\beta=2$ in the limit $N\rightarrow\infty$. The spacing produced by the gap have been ignored, see explanation in the main text.}
    \label{fig:SVspdistN2x}
\end{figure}

Thus, in Fig.~\ref{fig:SVspdistN2x} we display numerical results for the (unfolded) spacing distribution $P(s)$ among the particles in the gas corresponding to \eqref{jpdx} with the cubic potential plotted in Fig.~\ref{fig:potential} and for a large range of values of $g$, for matrices of size $N=32$. While the sizes of typical spacings are of order $1/N$ and fluctuating, for a density of particles like in Fig.~\ref{fig:SVdensity} which is supported on two disconnected intervals, there is one spacing, the one that goes over the gap, which is much larger (of order one) and more rigid than the other spacings. In a plot of the spacing distribution, the unfolding procedure would mix the spacing over the gap with the usual ones and produce artifacts. Therefore, in each sample, we disregard the spacing over the gap when we collect the spacings and calculate histograms from them.

The curve for $g=0.25$ (blue line) almost coincides with the $\beta=2$ Wigner-Dyson curve (dotted black line) corresponding to $g=0$. As $g$ increases the $P(s)$-curves (red curve for $g=250$, yellow curve for $g=2500$) deviate significantly from the Wigner-Dyson curve. Then, at the largest values of $g$ shown (purple curve for $g=2.5\times 10^4$ green curve for $g=2.5\times 10^5$), the curves tilt towards the Poisson curve (black dashed line), with clear tendency to an almost perfect fit with the latter as $Ng\rightarrow\infty$. 

\begin{figure}[h]
    \centering
    \includegraphics[width=1\linewidth]{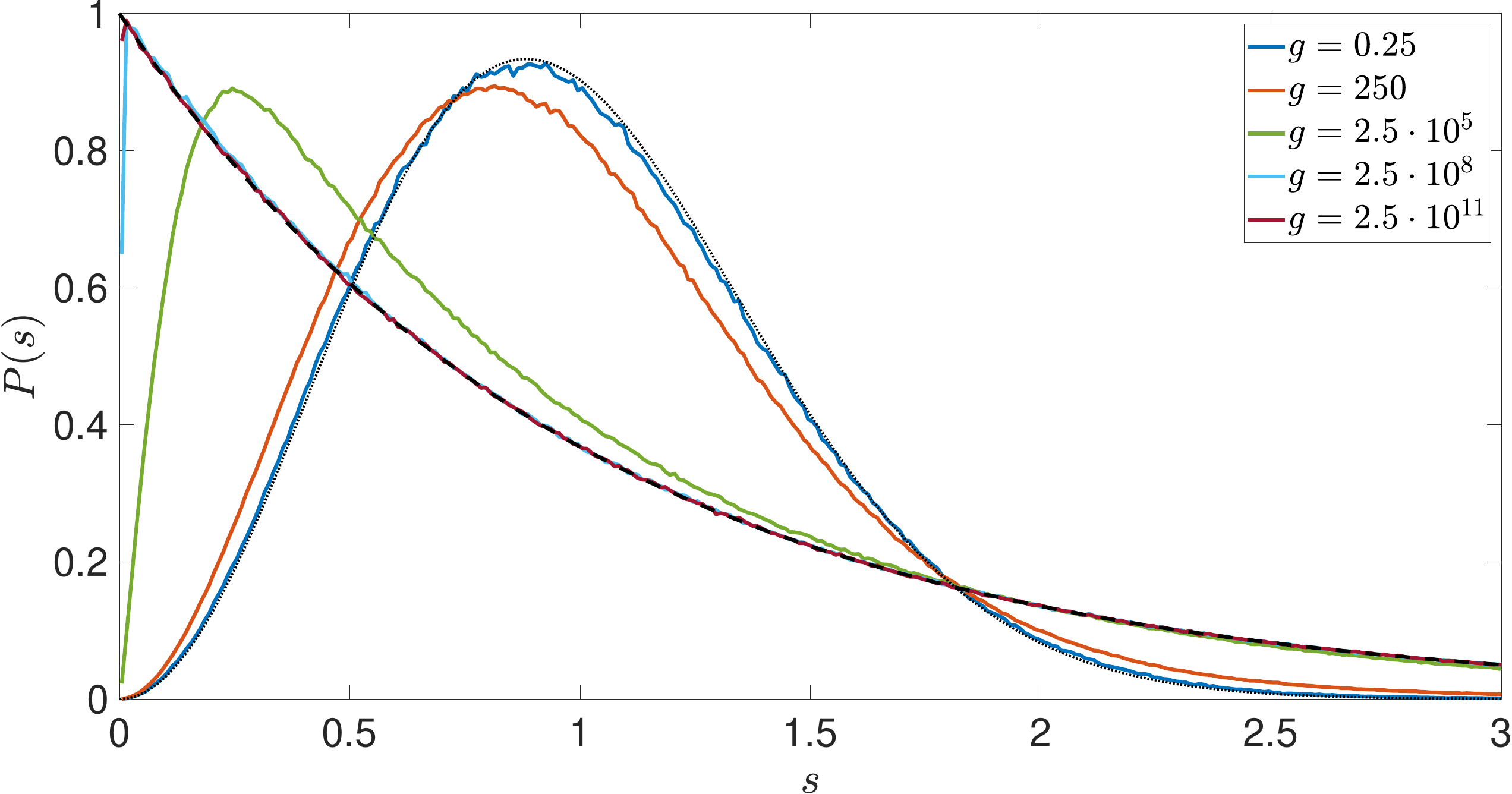}
    \caption{Similar plots as in Fig.~\ref{fig:SVspdistN2x} for the Ginibre potential $V(x)=x$ and for a larger range of $g$ values. }
    \label{fig:SVspdistN2vsN32}
\end{figure}

Fig.~\ref{fig:SVspdistN2vsN32} displays similar findings about $P(s)$ for Ginibre's case $V(x)=x$, for an even broader set of values of $g$. The behavior of the $P(s)$ curves is similar to Fig.~\ref{fig:SVspdistN2x} for the cubic potential, indicating the universality of repulsion as function of $g$. 

We refer the reader to Fig.~\ref{fig:SVspdistN2} in appendix \ref{AppC}, which displays the interparticle spacing distribution $P(s)$ for two particles (the analog of Wigner's surmise) for Ginibre's potential for comparison. 

To summarize, we have verified the same continuous crossover from Wigner-Dyson to Poisson statistics for $V=0$ (free fermions, Appendix \ref{AppB}), $V=x$ (Ginibre), $V=x^2$ (as in \cite{MNS}) and for the cubic \eqref{cubic1} potential, thus providing ample evidence for the universality of this phenomenon in the class of matrix models introduced and studied in this paper. 

\begin{figure}[h]
    \centering
    \includegraphics[width=1\linewidth]{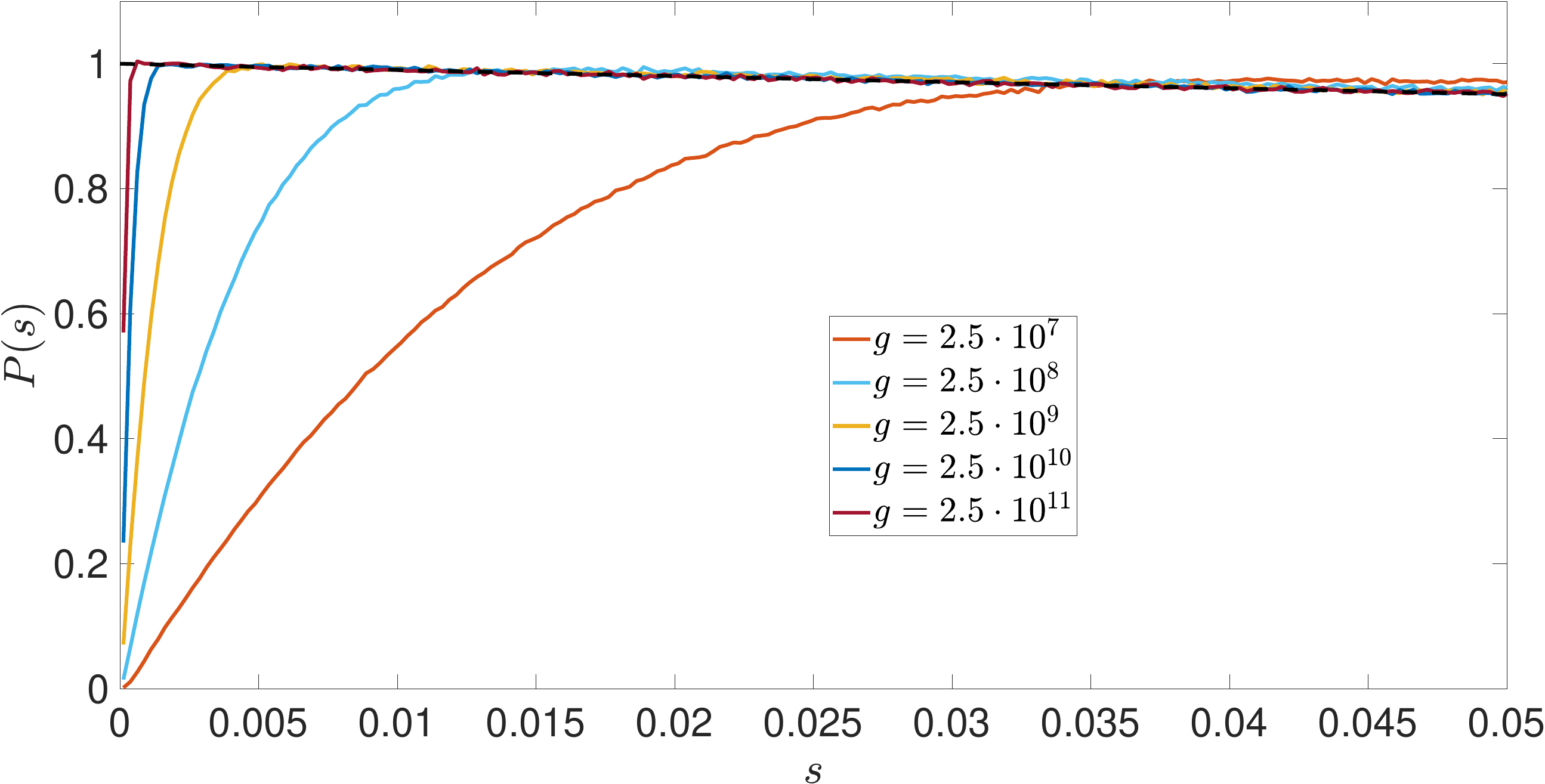}
    \caption{Magnification of the small-$s$ region in Fig.~\ref{fig:SVspdistN2vsN32} for the Ginibre potential for values of $g$ from $2.5\cdot 10^7$ to $2.5\cdot 10^{11}$. We use this plot to establish that $s_\text{micro}$ defined in the main text scales as $1/\sqrt{Ng}$.}
    \label{fig:SVspdistmag}
\end{figure}

We comment that the Poisson behavior at large values of $Ng$ kicks in at interparticle separation scales much larger than $s_{\rm micro}=1/\sqrt{Ng}$. This is so because the JPDF \eqref{jpdx} vanishes whenever two particles coincide in space and therefore $P(0)=0$. At large values of $Ng$, according to \eqref{jpdx}, $s_{\rm micro}$ must be the length scale below which the exact $P(s)$ departs from the Poisson curve and shoots down to zero, as can bee seen in Figs. \ref{fig:SVspdistN2x} and \ref{fig:SVspdistN2vsN32}. Thus, in summary, at large values of $Ng$, Poisson behavior holds universally on a range of length scales much larger than $s_{\rm micro}$. In order to demonstrate this ultra-short separation behavior of $P(s)$, we magnified this region in Fig.~\ref{fig:SVspdistN2vsN32} and displayed it in Fig.~\ref{fig:SVspdistmag}. 
We verified the scaling behavior $s_{\rm micro}=1/\sqrt{Ng}$ from Fig.~\ref{fig:SVspdistmag} by checking the $Ng$-dependence of various curves at half maximum where they plunge to zero. 

It is important to note that this change of statistics is a large-$Ng$ effect, and therefore cannot be related with the general phenomenon of breaking of the SRT, which seems to be a small-$g$ effect, as indicated in Fig.~\ref{fig:qualitative} (see also Fig.~\ref{fig:schematic}). It is only the large-$Ng$ end point of the flow, where the probability ensemble concentrates (as a probability measure) on normal matrices, that is correlated with Poisson statistics of the $x_i$.

\section{Singular Values as a Gas of non-interacting fermions on the positive half-line }\label{sec:Fermi}
In this section we show that the squares of singular values $x_i=\lambda_i^2$ can be interpreted as coordinates of non-interacting fermions living on the positive half-line.   

The simplest form of \eqref{jpdx} is obtained by setting $V(x)=0$. (Technically, in the absence of a confining potential, we should have the system confined in a big box or compactified on a large ring, in order to prevent the particles escaping to infinity, rendering \eqref{jpdx} normalizable.)

This case is discussed in detail in Appendix \ref{AppB}, where we also introduced some notations. 

As explained in Appendix \ref{AppB}, the determinantal form of \eqref{jpdx} (with $V(x)=0$) leads to its interpretation in terms of a gas of $N$ non-interacting free fermions occupying the positive half-line, in analogy with similar interpretation of the gas of eigenvalues of hermitian random matrices \cite{MNS, Parisi, BIPZ}, which typically occupies the whole real line. 

Restricting the particles to the positive half-line requires setting boundary conditions on their wave functions at the origin. As explained in Appendix \ref{AppB}, the gas consists of an even mixture of Dirichlet particles (with a single particle wave function obeying $\psi_D(0)=0$) and Neumann particles ($\psi_N'(0)=0$). 

This picture persists broadly also in the presence of a confining potential $V(x)$ in \eqref{jpdx}. Comparison of the integrand in \eqref{jpdx} with the analogous expression \eqref{Nfermion-diagonal} for free fermions (and the expressions \eqref{transfer} and \eqref{transfer-mixture} for free fermion propagators) leads us (upon identifying as usual the inverse temperature $\beta$ with Euclidean time) to define the full propagator of the $k$-th fermion
\begin{eqnarray}\label{Tk-pure}
\hat {\cal T}_k &\equiv& \left ( \begin{array}{cc} e^{-\beta \hat H_{D,k}} 
     & 0 \\
    0 & e^{-\beta \hat H_{N,k}} 
\end{array}\right)\nonumber\\{}\nonumber\\
&=& e^{-\frac{1}{2}NV(\hat x_k)} \hat {\cal T}_k^{(0)} e^{-\frac{1}{2}NV(\hat x_k)}
\end{eqnarray}
in a large graded Hilbert space with Dirichlet and Neumann subspaces, in the presence of the external potential $V(x)$. The operators $\hat x$ and $\hat p$ in \eqref{Tk-pure} are postulated to obey canonical commutation relations. Therefore, the hermitian Dirichlet and Neumann hamiltonians $\beta \hat H_{D,N;k}$ are well defined \cite{Parisi} by \eqref{Tk-pure}. 

As we show in Appendix \ref{AppB}, up to a normalizaton factor, one can obtain \eqref{Nfermion-diagonal} simply by considering the density matrix of the gas of $N$-free fermions on the {\em whole} real line (for which the single particle hamiltonian $\hat p^2$ is parity invariant), taking its diagonal matrix element in position space, and restricting coordinates of all particles to the positive half-line. This idea can be extended to work also in the presence of a confining potential $V(x)$. 

Thus, in order to construct the operators $\hat H_D$ and $\hat H_N$, we extend the original problem symmetrically onto the entire real line, with a potential term $V_s(\hat x)$ which is the {\em symmetric} extension of $V(\hat x)$ from the positive half-line onto the entire real line. (Namely, the operator version of $V(|x|)$.)
For this symmetric extension, we compute the matrix element 
\begin{equation}\label{Hs-prop}
\langle x|e^{-\beta \hat H_s}|y\rangle \equiv  \langle x|e^{-\frac{1}{2}NV_s(\hat x)} e^{-\beta \hat p^2} e^{-\frac{1}{2}NV_s(\hat x)}|y\rangle
\end{equation}
which defines the propagator of the parity-symmetric self-adjoint hamiltonian $\hat H_s$, whose eigenstates $\psi_n(x)$ are also eigenstates of the parity operator. Let us denote the corresponding eigenvalues $E_n$. The particles of the gas are expected to be confined, and we therefore assume the spectrum of $\hat H_s$ to be discrete, with real eigenstates. Let us take the spectral decomposition of \eqref{Hs-prop}
\begin{equation}\label{Hs-prop-spectral}
G_s(x,y;\beta) = \langle x|e^{-\beta \hat H_s}|y\rangle  = \sum_n e^{-\beta E_n} \psi_n(x)\psi_n(y)
\end{equation}
and split it into the contributions from parity-odd and parity-even states 
\begin{eqnarray}\label{even-odd-split}
\langle x|e^{-\beta \hat H_s}|y\rangle_{\rm odd}  &=& \sum_{\rm odd~states} e^{-\beta E_n} \psi_n(x)\psi_n(y)\nonumber\\{}\nonumber\\
\langle x|e^{-\beta \hat H_s}|y\rangle_{\rm even}  &=& \sum_{\rm even~states} e^{-\beta E_n} \psi_n(x)\psi_n(y)
\end{eqnarray}
Clearly, $\langle x|e^{-\beta \hat H_s}|y\rangle_{\rm odd}$ is separately odd in each of its arguments, whereas $\langle x|e^{-\beta \hat H_s}|y\rangle_{\rm even}$ is separately even in each of its arguments. Therefore, we can write
\begin{eqnarray}\label{even-odd-2}
\langle x|e^{-\beta \hat H_s}|y\rangle_{\rm odd}  &=& \frac{1}{2}\left(G_s(x,y;\beta) - G_s(x,-y;\beta)\right)\nonumber\\{}\nonumber\\
\langle x|e^{-\beta \hat H_s}|y\rangle_{\rm even}  &=& \frac{1}{2}\left(G_s(x,y;\beta) + G_s(x,-y;\beta)\right)\nonumber\\{}
\end{eqnarray}
in analogy with the free fermion case \eqref{DN1}. 

From the fact that $G_s(x,y;\beta)$ satisfies the time-dependent Schr\"odinger equation, with initial condition $\delta(x-y)$ at $\beta=0$, we see that the restriction of $\langle x|e^{-\beta \hat H_s}|y\rangle_{\rm odd}$ to the positive half-line also satisfies the Schr\"odinger equation with initial condition $\frac{1}{2}\delta (x-y)$, subjected in addition to Dirichlet boundary condition at the origin. We therefore identify it with the desired Dirichlet propagator 
\begin{equation}\label{Dirichlet-prop}
\langle x|e^{-\beta \hat H_D}|y\rangle = 2\langle x|e^{-\beta \hat H_s}|y\rangle_{\rm odd}\,,\quad x,y\geq 0\,.
\end{equation}
Similarly, the restriction of $\langle x|e^{-\beta \hat H_s}|y\rangle_{\rm even}$ to the positive half-line is the corresponding solution subjected to Neumann boundary condition at the origin, which we therefore identify with the desired Neumann propagator 
\begin{equation}\label{Neumann-prop}
\langle x|e^{-\beta \hat H_N}|y\rangle = 2\langle x|e^{-\beta \hat H_s}|y\rangle_{\rm even}\,,\quad x,y\geq 0\,.
\end{equation}
Thus, the parity-symmetric single-particle hamiltonian $\hat H_s$ for fermions living on the whole real line determines the Dirichlet and Neumann hamiltonians $\hat H_D$ and $\hat H_N$ and their propagators by means of \eqref{Dirichlet-prop} and \eqref{Neumann-prop}. This generalizes our discussion in Appendix \ref{AppB} of free fermions, with $\hat H_s=\hat p^2$ and the associated propagators \eqref{DN1}. 

Note that this construction of the Dirichlet and Neumann propagators justifies the tacit assumption made in \eqref{Tk-pure} that the inverse temperature $\beta$, which depends on $g, N$ and the other couplings in $V(x)$, must be common to both the Dirichlet and Neumann sectors.

The quasi-propagator of the mixed Dirichlet-Neumann fermion, analogous to \eqref{transfer-mixture}, whose matrix element appears inside the determinant in \eqref{jpdx}, is of course 
\begin{equation}\label{Tk-mixture}
\hat T_k = \frac{1}{2}{\rm tr} \hat {\cal T}_k = e^{-\frac{1}{2}NV(\hat x_k)} \hat T_k^{(0)} e^{-\frac{1}{2}NV(\hat x_k)}\,.    
\end{equation}
Thus, it follows from \eqref{Dirichlet-prop} and \eqref{Neumann-prop} that 
\begin{equation}\label{Tk-mixture-xy}
\langle x|\hat T|y\rangle = G_s(x,y;\beta) = \langle x|e^{-\beta \hat H_s}|y\rangle \,,\quad x,y\geq 0\,.
\end{equation}
That is, the fermion quasi-propagator on the half-line is nothing but the value of the proper propagator \eqref{Hs-prop-spectral} of the operator $\hat H_s$, with both arguments restricted to the positive half-line. This is of course not a proper propagator, since \eqref{Tk-mixture-xy} does not have the semi-group property on the positive half-line. This by no means leads to any contradiction, since on the positive half-line \eqref{Hs-prop-spectral} is not an expansion in terms of orthogonal functions: Only eigenstates $\psi_n(x)$ which on the full line have the same parity are also orthogonal on the positive half-line, but not eigenstates of opposite parity. (For example, the even ground state $\psi_0(x)$ and the first excited odd state $\psi_1(x)$ can be both chosen positive on the positive half line.) On the other hand, and for the same reason, the restriction of both expressions in \eqref{Dirichlet-prop} and \eqref{Neumann-prop} to the positive half-line remains an expansion in terms of orthogonal functions, in accordance with the fact that both expressions are proper propagators on the positive half-line. 

To summarize, in principle, one has to derive from \eqref{Tk-pure} the parity-even hamiltonian $\hat H_s$ which governs particles living on the whole real line, find its eigenstates (both even and odd) and propagator \eqref{Hs-prop} (or \eqref{Hs-prop-spectral}). We then identify the kernel inside the determinant in \eqref{jpdx} as the single-particle quasi-propagator $\hat T_k$ in \eqref{Tk-mixture-xy}, and following \cite{Parisi, MNS, BIPZ} and the discussion in Appendix \ref{AppB} for free fermions, we interpret the integrand of \eqref{jpdx} as the diagonal matrix element (here $|\Vec{x}\rangle_A$, defined in \eqref{Nketasym}, is the antisymmetrized position eigenstate of $N$ particles)
\begin{equation}\label{Nfermion-density}
   \rho(x_1,x_2,\ldots,x_N) = {}_A\langle \Vec{x}|\hat T_1 \hat T_2 \cdots \hat T_N|\Vec{x}\rangle_A
\end{equation}
of the canonical-ensemble density matrix of a gas of non-interacting fermions confined to the positive half-line, thus generalizing the analogous expression \eqref{Nfermion-diagonal} for free fermions. Once $\hat H_s$ and its eigenstates are known explicitly, one can apply the methods of quantum statistical mechanics to compute the statistics of the gas occupying the whole line. 
The statistical properties of the gas of singular values in \eqref{jpdx}, confined to the positive half-line, can then be deduced simply by focusing on the right half of the gas governed by $\hat H_s$, after proper rescaling of its density by a factor of $2$, coming from the need to normalize each eigenstate $\psi_n(x)$ in \eqref{Hs-prop-spectral} on the half-line rather than on the full line. 

For a generic matrix potential $V(x)$, and in particular, for the cubic (in $\phi^\dagger\phi$) potential \eqref{cubic1} analyzed specifically in the present paper, it is practically impossible to obtain $\beta \hat H_s$ (and therefore $\beta \hat H_{D,N}$) in closed form. 

A rare exception is the harmonic potential $V(x) = x^2$ mentioned in the previous section. In this case we can write the matrix element inside the determinant in \eqref{jpdx} as  
$e^{-N(g+\frac{1}{2})(x_i^2 + x_j^2) + 2Ngx_i x_j}$, which is readily recognized, {\em for a problem defined on the whole real line} \cite{MNS}, as being proportional to the matrix element of the Euclidean-time propagator of an harmonic oscillator
\begin{eqnarray}\label{HO}
&&\langle x_i| e^{-\beta H_s}|x_j\rangle = \langle x_i| \exp\left[-\beta\left(\frac{\hat p^2}{2m} + \frac{m\omega^2 \hat x^2}{2}\right)\right]|x_j\rangle = \nonumber\\
&&C\exp\left[-\frac{m\omega}{2\hbar}\left(\coth (\beta\hbar\omega)\,(x_i^2+x_j^2) -\frac{2x_ix_j}{\sinh (\beta\hbar\omega)}\right)\right]\nonumber\\{}
\end{eqnarray}
(where $C=\sqrt{\frac{m\omega}{2\pi\hbar\sinh(\beta\hbar\omega)}}$), whose mass $m$, frequency $\omega$ and inverse temperature $\beta$ are related by 
\begin{equation}\label{beta-omega}
    \cosh (\beta\hbar\omega) = 1 + \frac{1}{2g}
\end{equation}
and 
\begin{equation}\label{mass-omega}
    \frac{m\omega}{\hbar} =N\sqrt{1+4g}
\end{equation}
(the latter being the inverse length unit squared). Thus, eigenvalue statistics in \cite{MNS} is reduced to the canonical ensemble of a gas of non-interacting fermionic harmonic oscillators occupying the whole real line, rendering this model exactly solvable. 

The Dirichlet and Neumann components of \eqref{HO} are  
\begin{eqnarray}\label{HO-Dir}
&&\langle x_i| e^{-\beta \hat{\cal H}_D}|x_j\rangle = 2C\exp\left[-\frac{m\omega}{2\hbar}\coth (\beta\hbar\omega)\,(x_i^2+x_j^2)\right]\nonumber\\
&&\cdot\sinh \left(\frac{m\omega x_ix_j}{\hbar\sinh (\beta\hbar\omega)}\right)\,,
\end{eqnarray}
and
\begin{eqnarray}\label{HO-Neum}
&&\langle x_i| e^{-\beta \hat{\cal H}_N}|x_j\rangle = 2C\exp\left[-\frac{m\omega}{2\hbar}\coth (\beta\hbar\omega)\,(x_i^2+x_j^2)\right]\nonumber\\
&&\cdot\cosh \left(\frac{m\omega x_ix_j}{\hbar\sinh (\beta\hbar\omega)}\right)\,,
\end{eqnarray}
and the quasi-propagator $\langle x_i|\hat T|x_j\rangle$ in \eqref{Tk-mixture-xy} is given simply by the restriction of \eqref{HO} to the positive half-line. Following \cite{MNS}, one then computes the statistical properties of the Fermi gas in the canonical ensemble, from which we can compute the statistics of singular values in \eqref{jpdx} by focusing on the right-hand side of the Fermi gas, and rescaling the overall density by a factor of 2.

The $V(x)=x^2$ model in \cite{MNS} is exactly solvable. In particular, the crossover from Wigner-Dyson (at small $g$) to Poisson statistics (at large $g$) was established there analytically. In our case, it corresponds to flowing to normality for the purely quartic matrix potential $(\phi^\dagger\phi)^2$, in which case (for $g=0$) the singular values always condense in a single segment touching the origin, the support of complex eigenvalues of $\phi$ is a disk \cite{feinberg1997non}, and therefore does not teach us anything about the SRT. However, this case serves as an exactly solvable analytical handle for understanding the behavior of this type of models for general potentials $V(x)$. In particular, the Vandermode-determinant-squared in the denominator of the Itzykson-Zuber integral \eqref{eq:IZ} knocking out the similar factor in the measure in \eqref{partition2} is in complete analogy with what happens in \cite{MNS}. (This cancellation is also analogous to the cancellation of pairs of Vandermonde determinants in the numerator with a pair in the denominator at intermediate points along the time evolution of the path integral formulation \cite{CMM, BKZ} of the quantum mechanical matrix model in \cite{BIPZ}.)

Quite surprisingly, an explicit construction of $\hat H_s$ (and $\beta \hat H_{D,N}$) as in \eqref{HO} is not available for the simpler Ginibre model, $V(x)=x$. In this case, the matrix element inside the determinant in \eqref{jpdx} is $e^{-Ng(x_i-x_j)^2 -\frac{1}{2}(x_i + x_j)}$, which is proportional to the matrix element of the Euclidean-time propagator of a particle under the influence of a linear potential $V(x)=x$ along the {\em entire} real line \cite{FH, Schulman}. However, our problem is defined on the positive half-line with potential $V(x)=x$, and the extended symmetric problem on the entire real line is for the potential $V(x)=|x|$, for which no explicit closed form for the propagator is known. 

Save for free fermions (as discussed in Appendix \ref{AppB}) and the very special case $V(x)=x^2$ analyzed here (following \cite{MNS}), explicit analytic analysis of the Fermi gas of singular values $x_i$ is not available for a generic potential $V(x)$. 
However, following the discussion in the previous section, it is clear that the continuous change in statistics of singular values from $\beta=2$-repulsive Wigner-Dyson statistics to non-repulsive Poisson statistics, established by \cite{MNS} for $V(x)=x^2$, should be universal and hold for general matrix potentials $V(x)$. This is obvious for the two endpoints of the flow. The squared Vandermonde determinant in \eqref{partition2} implies Wigner-Dyson statistics at $g=0$ for any $V(x)$. At the other endpoint as $Ng\rightarrow\infty$, corresponding to normal matrices, repulsion is suppressed (on interparticle separation scales larger than $1/\sqrt{Ng}$), leading to Poisson statistics of interparticle spacings. In between, the flow is continuous in $g$. In Fig.~\ref{fig:SVspdistN2x} in the previous section we have provided numerical verification of this conjecture for the cubic potential \eqref{cubic1}, and in Fig.~\ref{fig:SVspdistN2vsN32} we provide such numerical verification for Ginibre's potential $V(x)=x$. Moreover, the basic properties of diffusion and the behavior of the heat kernel at finite and small Euclidean time $\tau$ establish this conjecture for free particles at the end of Appendix \ref{AppB}. 

 With all that said, we stress that this plausibly universal continuous change of statistics has nothing to do with breaking of the SRT, which happens at small values of $g$. (Recall Fig.~\ref{fig:schematic}.)

Given the lack of explicit analytical description of the gas \eqref{jpdx} in case of a generic potential $V(x)$, for the more immediate goal of computing the density of singular values as it changes along the flow for a generic $V(x)$, we might as well analyze \eqref{jpdx} numerically, as we did for the cubic potential \eqref{cubic1}. Such analysis would be naturally cheaper in computation time than that of simulating the full matrix model. However, reconstruction of the average spectrum of eigenvalues of the non-hermitian matrix $\phi$ from that of its singular values at some finite value of $g$ along the flow is tricky. We shall have to say more on this problem in Section~\ref{sec:conjecture}. 

\subsection{Compressibility of the Gas of Singular Values}\label{subsec:compressibility}

The Fermi gas \eqref{jpdx} of singular values is in thermal equilibrium at some temperature $T$, which must be a monotonically increasing function of the coupling $g$. This is explicit in the free fermion case discussed in Appendix \ref{AppB}, where in fact $T\propto g$ as is evident in \eqref{Ng} ($\tau$ is the Euclidean time). In the quadratic case $V(x) = x^2$, Eq.\eqref{beta-omega} provides yet another example of such a monotonic relation (for fixed $\omega$). Furthermore, in the large-$g$ limit, we have seen that the many-body position-space diagonal matrix element of the density matrix \eqref{jpdx} of the Fermi gas tends asymptotically to the i.i.d. distribution \eqref{jpdxiid}, namely, the classical Boltzmann distribution, which describes our Fermi gas in the high temperature limit. Large $g$ means large $T$ for any confining potential $V(x)$. In this limit the average density \eqref{rhox} tends simply to 
\begin{equation}\label{rhox1}
  \rho(x)\propto e^{-NV(x)}  
\end{equation}
meaning the singular values, or fermions, condense in the minima of $V(x)$.

To summarize, under the flow of the matrix model \eqref{prob1} with increasing $g$ from the Wigner-Dyson point \eqref{prob} at $g=0$ to the normal matrix point as $g\rightarrow\infty$, the Fermi gas \eqref{jpdx} heats up from $T=0$ to its classical Boltzmann limit \eqref{jpdxiid} as $T\rightarrow\infty$. 

As the Fermi gas heats up, its thermodynamic properties obviously change with $T$. In particular, its compressibility {\em increases} (i.e., it becomes easier to compress), as we now demonstrate numerically. We hasten to remark that our discussion of compressibility is rather intuitive, based on the expected qualitative thermodynamic bahavior of the Fermi gas with increasing temperature. Rigorous analysis of the analogous issue, namely, spectral rigidity, and its relation to Anderson localization in the matrix model \cite{MNS} was reported in \cite{Kravtsov1, Kravtsov2}, based on the general theory developed in \cite{Chalker1996}. Applying the formalism of \cite{Chalker1996} rigorously to our model deserves a separate effort.

The interparticle Pauli repulsion associated with \eqref{jpdx} for $g>0$ is short-ranged. For large enough $Ng$, the interparticle repulsion length scale is clearly $1/\sqrt{Ng}$. In contrast, when $g=0$, interparticle repulsion is long-ranged, due to the unscreened Coulomb interaction (resulting from the Vandermonde factors in \eqref{partition2}). In the latter case, every particle is repelled strongly by all the other particles. As $g$ increases along the flow, interparticle repulsion gets progressively suppressed. Therefore, the gas of singular values at $g=0$ should be less compressible than when $g>0$. Compressibility of our gas should be an increasing function of $g$ (and therefore of $T$). 

\begin{figure}[h]
    \centering
    \includegraphics[width=1\linewidth]{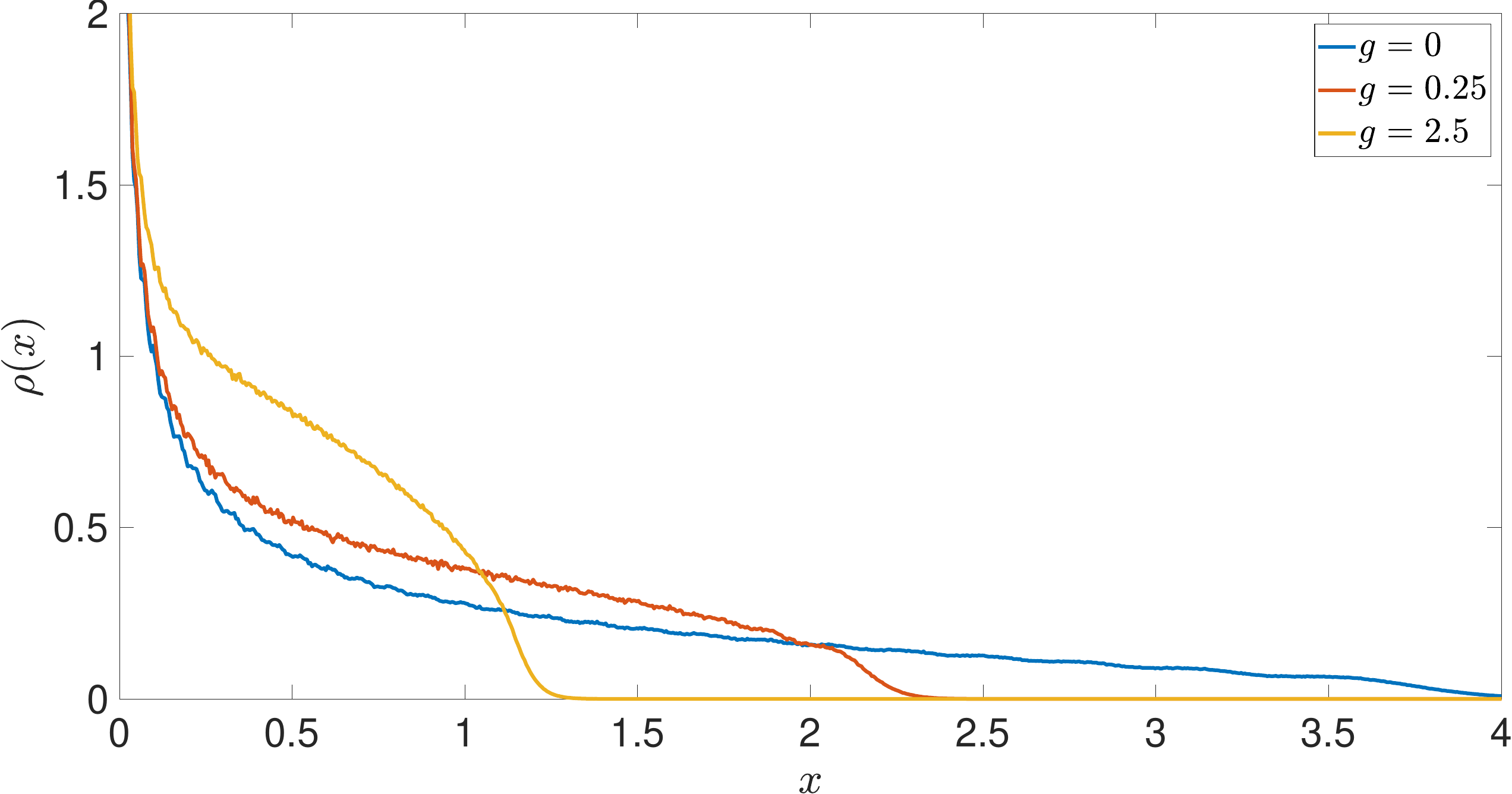}
    \caption{Solid curves show the density of square singular values for Ginibre's potential $V(x)=x$ and for a few values of $g$. It was obtained from Monte Carlo simulation for $N=32$ singular values. The support of the density of particles shinks as $g$ increases, while keeping the confining potential (the applied pressure) fixed.}
    \label{fig:SVdensityGin}
\end{figure}

We checked numerically that this is indeed the case for Ginibre's model with $V(x)=x$. In Fig.~\ref{fig:SVdensityGin} we plot the density of singular values of this gas for $g=0,0.25$ and $2.5$, for matrices of size $N=32$. 
At $g=0$, the squared singular values of Ginibre's model condense in the largest segment shown $0\leq x\lesssim 4$. We now increase $g$ while keeping the confining potential $V(x)=x$ fixed. In other words, we reduce interparticle repulsion under fixed external pressure. As expected, the result is that the $x_i$ get squeezed into smaller condensation segments, which decrease as $g$ increases. 

The increased compressibility of the gas as $g$ (and therefore temperature) increases is just another facet of the decreasing level-repulsion in the original model \cite{MNS}. As was mentioned in the previous section, this feature of that model was the original motivation of the authors of \cite{MNS}, who introduced this model of a disordered (hermitian) system with continuous crossover from the ergodic phase of extended states (with $\beta=2$ Wigner-Dyson level statistics) to a phase of localized states, with a preferred basis and Poisson level statistics.

\section{Approximate Reconstruction of the Distribution of Complex Eigenvalues from that of the Singular Values and a related canonical ensemble of random permutations}\label{sec:conjecture}
In contrast with the two previous sections, the discussion in this section is somewhat heuristic and conjectural, and also lies off the main theme of the paper about breaking of the SRT and crossover in the spacing statistics along the flow with the parameter $g$. 

The JPDF of singular values $\Pi(\lambda_1,\dots,\lambda_N)$ is given at each point along to flow by \eqref{jpdsv} as function of the flow parameter $g$. Can we deduce from \eqref{jpdsv} anything useful about the average distribution of eigenvalues $z_i$ of the non-hermitian matrices $\phi$ at some arbitrary $g$?

The answer to this question is known at both endpoints $g=0$ and $g\rightarrow\infty$ of the flow:

At $g=0$, one of the main results of \cite{feinberg1997non}, which lead to the SRT, is the equation
\begin{equation}\label{Fgamma}
    \gamma\left[ r^2F\left(\frac{\gamma r^2}{\gamma -1}\right)-\gamma +1\right] =0
\end{equation}
relating the Cauchy transform of the density of the $x_i=\lambda_i^2$, namely, the resolvent $F(w)$ of $\phi^\dagger\phi$, and the integrated radial density of eigenvalues $\gamma (r)$ of $\phi$.

At the other endpoint $g\rightarrow\infty$, namely, the endpoint of normal matrices, as explained in Appendix \ref{AppA}, we simply have $z_i=\lambda_i e^{i\theta_i}$, with the $\theta_i$ statistically independent of the $\lambda_i$ and uniformly distributed in $[0,2\pi)$. (The resulting ensemble is not to be confused, of course, with the canonical ensemble \eqref{zpdf} of normal matrices.) 

For an arbitrary value of $g$, the singular values $\Lambda$ and unitary matrices $W$ are statistically correlated in a non-trivial way, and we cannot just draw them independently from their corresponding marginal distributions (namely, \eqref{jpdsv} for $\Lambda$, and the distribution for $W$ obtained from \eqref{partition2} after integrating over $\Lambda$, which we cannot compute explicitly in closed form for a general potential $V$ anyway).  

Here we suggest a heuristic non-rigorous approach for constructing the statistical ensemble of non-hermitian matrices $\phi$ along the flow, given a quenched (or self-averaged) set of SVs $\Lambda$, which we expect to be a good approximation in the large-$N$ limit. 

In passing from the probability distribution \eqref{prob1} for $\phi$ to the JPDF \eqref{jpdsv} of singular values, we integrated over the unitary matrix $W=VU$ to obtain the Itzykson-Zuber result \eqref{eq:IZ}. It is well known \cite {Szabo, Szabo-rev} that the saddle point approximation for the Itzykson-Zuber integral over the unitary group is actually exact for any value of $N>1$, a result which is a special case of the Duistermaat-Heckman theorem. There are $N!$ saddle points of the integral on the LHS of \eqref{eq:IZ} at each of which 
\begin{equation}\label{saddle}
    W_P = DP
\end{equation}
where $D\in U(1)^N$ is a diagonal unitary matrix and $P$ is the unitary matrix
\begin{equation}\label{P}
    P_{ij} = \delta_{i,P(j)} = \delta_{P^{-1}(i),j}
\end{equation}
corresponding to the permutation $P\in S_N$. One then computes the contribution of Gaussian fluctuations around the saddle point $W=W_P e^{i\epsilon}$ by integrating over the hermitian matrix $\epsilon$, and after summing over the contribution of {\em all} $N!$ saddle points, one obtains the RHS of \eqref{eq:IZ}. 

The saddle points $W_P$ dominate the integral \eqref{eq:IZ} locally in the unitary group, and each such saddle induces correlation between the unitary matrices $U$ and $V$ in \eqref{SVD}, namely
\begin{equation}
    V=DPU^\dagger
\end{equation}
and therefore picks up a particular non-hermitian matrix 
\begin{equation}\label{phiP}
    \phi_p=U\Lambda V = U(\Lambda DP) U^\dagger
\end{equation}
together with its adjoint $\phi_P^\dagger  = U(P^\dagger D^\dagger \Lambda) U^\dagger$, which is an extremal point of the ensemble \eqref{prob1} (as far as the unitary degrees of freedom are concerned). The diagonal matrix 
\begin{equation}\label{zeta}
\zeta=\Lambda D
\end{equation}
in \eqref{phiP} has complex entries whose moduli are determined by the $\lambda_i$, with uniformly distributed random phases. It is multiplied by the unitary permutation matrix $P$, which has only a single non-zero element $1$ in each row and each column. Thus, the non-hermitian matrix 
\begin{equation}\label{psiP}
    \psi_P = \zeta P
\end{equation} 
in \eqref{phiP} (which plays the same role as the fiducial matrix $\phi_f$ in \eqref{fiducial}) has just one non-vanishing complex entry in each row and each column. Since $U$ is distributed uniformly, we see that the dominant $\phi_P$ in \eqref{phiP} is actually the orbit of $\psi_P$ through $U(N)$.

To get an idea of the eigenvalue spectra of the matrices $\psi_P$, consider for example the case $N=3$. The six permutation matrices are the identity $P_e={\bf 1}$ and the two cyclic permutations $P_1$ and $P_2$ (altogether, the three even permutations forming the alternating group $A_3$, which in this case is just the cyclic group of order 3), and the three pair interchanges $P_{(12)}=P_3, P_{(23)}=P_4$ and $P_{(13)}=P_5$ (the three odd permutations). We have of course $\psi_e={\rm diag} (\zeta_1,\zeta_2,\zeta_3)$. One can then readily compute the remaining five matrices and diagonalize them. The results are that $\psi_{P_1}$ and $\psi_{P_2}$ are isospectral, with eigenvalues $(\zeta_1\zeta_2\zeta_3)^{1/3}, (\zeta_1\zeta_2\zeta_3)^{1/3} w$ and $(\zeta_1\zeta_2\zeta_3)^{1/3} w^2$, where $w=e^{2\pi i/3}$. These eigenvalues are completely symmetric functions of the three $\zeta_i$. The eigenvalues of each of the remaining three $\psi_P$'s, which correspond to a pair interchange, are the single $\zeta_i$ unaffected by the interchange, together with $\pm$ the geometric mean of the two interchanged $\zeta_i$'s. Thus, for example, the eigenvalues of $\psi_{(12)}=\psi_{P_3}$ are $\pm\sqrt{\zeta_1\zeta_2}$ and $\zeta_3$. To summarize this example, given $P$, we see that eigenvalues of $\psi_P$ associated with a cycle of order $n$ in $P$ which mixes a subset of $n$ $\zeta_i$'s, is the geometric mean of these $\zeta_i$'s multiplied by the $n$ roots of unity of order $n$ ($n=1$ for $P_e$, $n=1,2$ for each of the odd permutations, and $n=3$ for the two cyclic ones). $P_1$ and $P_2$ are each full cycle of order 3 and therefore isospectral. This pattern should obviously persist for general values of $N$. Thus, let $P\in S_N$ be broken into its cycles. A particular cycle of length $n$ mixes $n$ out of the $N$ original $\zeta_i$'s (each $\zeta_i$ is affected by only one cycle, of course), and leads to $n$ eigenvalues of $\psi_P$ given by the geometric mean of the $\zeta_i$'s involved, multiplied by $e^{2\pi i k/n}\,, k=0,1,\ldots n-1$. The full eigenvalue spectrum of $\psi_P$ is the union of all these $n$-tuples over all cycles.

Let us now return to the main discussion. From \eqref{phiP} we obtain $\phi_P\phi_P^\dagger = U\Lambda^2 U^\dagger$ and $\phi_P^\dagger\phi_P = U P^\dagger \Lambda^2P U^\dagger$. It is useful to define the permuted array of singular values
\begin{equation}\label{lambdaP}
    \Lambda_P = P^\dagger \Lambda P = {\rm diag}(\lambda_{P(1)}, \lambda_{P(2)},\ldots,\lambda_{P(N)})
\end{equation}
so that $\phi_P^\dagger\phi_P = U \Lambda_P^2 U^\dagger$. Thus, $[\phi_P,\phi_P^\dagger] = U(\Lambda^2-\Lambda_P^2)U^\dagger$ and therefore at the $P$-th saddle point of \eqref{eq:IZ}, the positive-definite commutator-squared term penalizing for deviation from normality is 
\begin{eqnarray}\label{penalizer}
    C(P;\Lambda) &=& {\rm Tr}\left([\phi_P,\phi_P^\dagger]^2\right)\nonumber\\ 
    &=&  {\rm Tr}\left(\Lambda^2-\Lambda_P^2\right)^2 = \sum_{i=1}^N (\lambda_i^2-\lambda_{P(i)}^2)^2\,.
\end{eqnarray}
It is easy to see that $C(P;\Lambda)=C(P^{-1};\Lambda)$.

As was mentioned above, in order to obtain the Itzykson-Zuber result \eqref{eq:IZ} from the saddle point approximation, we have to compute the Gaussian fluctuations at each saddle point, and sum over contributions of all $N!$ saddle points. Here we propose to just take the leading large-$N$ approximation, that is, evaluate the LHS of \eqref{eq:IZ} at each saddle point $W_P$ (without the fluctuations around it), and sum over all $N!$ saddle points. 

Henceforth we shall compute $C(P;\Lambda)$ for a fixed, judicially chosen quenched $\Lambda$, which has to be non-degenerate. This is so because in effect we are approximating the integral over $W$ in \eqref{partition2}, and the probability to draw a degenerate $\Lambda$ from the ensemble associated with \eqref{partition2} is null. With no loss of generality, we shall also assume these entries are increasing in order. In fact, since our goal is to approximate \eqref{eq:IZ} by summing over the dominant saddle points, we might as well assume that the entries of $\Lambda$ are distributed according to \eqref{jpdsv}, which is the result of exact integration over $W$ in \eqref{partition2}. The JPDF \eqref{jpdsv} does exhibit level repulsion for any $g\geq 0$ due to its determinantal structure. Thus, it makes sense to pick our quenched $\Lambda$ as the most probable self-averaging (and $g$-dependent) configuration $\Lambda_g = {\rm diag}(\lambda_{g1}, \lambda_{g2},\ldots,\lambda_{gN})$ drawn from \eqref{jpdx}, which leads in the large-$N$ limit to the finitely supported smooth density discussed in the previous section. 

Given this quenched $\Lambda_g$ configuration, which dominates the partition function \eqref{partition}, we approximate the integral representation \eqref{partition2} for this partition function in two steps. First, we approximate the integration over the SVs by the value of integrand at our dominant $\Lambda_g$ configuration. Second, for this fixed $\Lambda_g$, we then compute the leading saddle point approximation of the integral over $W$ in \eqref{eq:IZ} by summing over all saddle points $W_P$ in \eqref{saddle} (not including Gaussian fluctuations around each saddle point)
\begin{eqnarray}\label{eq:IZapprox}
&&\int_{U(N)} d\mu(W) e^{2Ng{\rm Tr}(\Lambda_g^2 W^\dagger\Lambda_g^2 W)}
\simeq\nonumber\\{}\nonumber\\ &&\sum_{P\in S_N} e^{2Ng{\rm Tr}(\Lambda_g^2 P^\dagger\Lambda_g^2 P)} = \sum_{P\in S_N} e^{2Ng{\rm Tr}(\Lambda_g^2\Lambda^2_{gP})}    
\end{eqnarray}
where in the last step we have used \eqref{lambdaP}, and feed this result into our approximation to \eqref{partition2} in the first step. We end up with 
\begin{eqnarray}\label{partition4}
&&{\mathcal{Z}_N}\simeq
{\rm Const.}(\prod_k \lambda_{gk})\,\Delta^2(\Lambda_g^2)\, e^{-N\sum_iV(\lambda_{gi}^2) }\cdot \nonumber\\{}\nonumber\\&&\sum_{P\in S_N}e^{-Ng C(P;\Lambda_g)}
\end{eqnarray}
where we have used \eqref{penalizer}. Thus, given our quenched $\Lambda_g$ configuration, this approximate expression for $\mathcal{Z}_N$ induces a probability distribution over the $N!$ permutations 
\begin{eqnarray}\label{permutations}
{\cal P}(P|\Lambda_g) &=& \frac{e^{-Ng C(P;\Lambda_g)}}{\sum_{Q\in S_N}e^{-Ng C(Q;\Lambda_g)}}\\ \nonumber\\
&=& \frac{e^{-Ng C(P;\Lambda_g)}}{{\rm Perm}_{ij}\left(e^{-Ng (x_{gi}-x_{gj})^2}\right)}\,,
\end{eqnarray}
where the denominator in the last expression is the permanent of the indicated matrix. 

The next step in our approximation is to construct an ensemble of non-hermitian matrices, all sharing the same prescribed set of singular values $\Lambda_g$ (up to a permutation), in the most natural way, namely, construct the matrix $\phi_P$ according to \eqref{phiP}, with $P$, $U$ and $D$ drawn independently from their corresponding ensembles. $P$ is drawn from \eqref{permutations}, $U$ is Haar-distributed over $U(N)$, and $D$ is uniformly distributed over $U(1)^N$. This defines uniquely the desired ensemble of non-hermitian matrices. By construction, all matrices in this ensemble share the same set of singular values $\Lambda_g$, which was chosen to be the dominant self-averaging configuration of \eqref{jpdx}. Therefore, we conjecture that in the large-$N$ limit, this ensemble would lead to the same average density of complex eigenvalues as \eqref{prob1}, at the corresponding value of $g$.

Note that if we replace in \eqref{permutations} the quenched, self-averaging configuration $\Lambda_g$ by some arbitrary $g$-independent array $\Lambda$ of non-degenerate singular values, we obtain yet another ensemble of random permutations, unrelated to our matrix ensemble \eqref{partition1} and the Fermi gas \eqref{jpdx} associated with it
\begin{equation}\label{permutations1}
{\cal P}(P|\Lambda) = \frac{e^{-Ng C(P;\Lambda)}}{\sum_{Q\in S_N}e^{-Ng C(Q;\Lambda)}}\,.
\end{equation}
In this canonical ensemble, $Ng$ plays the role of {\em inverse temperature}. This is in contrast to \eqref{permutations} and \eqref{jpdx}, in which temperature is some monotonically increasing function of $g$, as was discussed in Section \hyperref[subsec:compressibility]{3.1}. 

At $g=0$ (the high temperature limit) the ensemble \eqref{permutations1} is maximally random, that is, completely unbiased - all permutations are equiprobable. At the other endpoint as $g\rightarrow\infty$ (zero temperature), any $C(P;\Lambda)>0$ has zero probability. Only $C(e;\Lambda)=0$ survives, and ${\cal P}(P|\Lambda) = \delta_{P,e}$ localizes on the identity permutation. 

The width of the distribution of permutations shrinks from its maximal value $N!$ at $g=0$ to just $1$ as $g\rightarrow\infty$. It is thus natural to wonder whether the width of the distribution \eqref{permutations1} is a monotonically decreasing function of $g$. Moreover, for a given $g>0$ and $\Lambda$, is \eqref{permutations1} dominated by a single permutation $P^*$? If so, how narrowly is it peaked around $P^*$? These are by no means simple questions, since a fixed given value $C_0$ of the penalty term \eqref{penalizer} may be realized by just a small number of permutations with large cycles, or by very many permutations with short cycles. That is, entropy is an issue to reckon with in coping with these questions. Another interesting question is whether \eqref{permutations1}, with its astronomically large number of states (in the large-$N$ limit), and dependence on the `external field' $\Lambda$, can be mapped on one of the conventional statistical mechanical models. (In this respect, the ensemble \eqref{permutations} is much more complicated than \eqref{permutations1}, because $\Lambda_g$ varies along the $g$-flow.) 

These questions clearly indicate that the probability ensemble \eqref{permutations1} is interesting on its own as a problem in the general theory of probability over the symmetric group $S_N$. This is so especially in the large-$N$ limit, where some analytical tools are known about asymptotically large permutations\cite{Vershik, Vershik1, Logan}.

\section{Numerical Methods}\label{sec:methods}

Our numerical simulations focus on the cubic potential \eqref{cubic1} for the reasons mentioned following that equation. For concreteness we set $m^2=16$, $\alpha=-16$, $u=3$ in \eqref{cubic1} (see Fig.~\ref{fig:potential}) and simulate the ensemble \eqref{prob1} for various values of the flow parameter $g$. Our aim is to verify numerically the scenario described in the Introduction for breaking of the SRT and to estimate the critical value $g_{\rm crit}$ at which the SRT breaks for the first time.

To this end, we have run Monte Carlo simulations in which we sequentially propose a random move, as described below, and accept or reject it according to the standard Metropolis algorithm\cite{abraham1986computational,bhanot1988metropolis,chib1995understanding,krauth2006statistical}, thereby obeying detailed balance, which is the necessary condition for convergence to the desired JPDF. We have run two different kinds of Monte Carlo simulations: One for the squared singular values (SSV), following the JPDF \eqref{jpdx} (or the corresponding JPDF derived from \eqref{partition2} for $g=0$), and the other for full matrices, following the probability distribution \eqref{prob1}. The former has the advantage that it consists of only $N$ real positive random variables, while the latter involves $N^2$ complex entries. Since in \eqref{jpdx} and \eqref{partition2} the eigenvectors are integrated out, the former can only measure statistics of the SSV, whereas from the full-matrix simulation we can obtain any statistics, for example the ensemble average of $\mathrm{Tr}[\phi,\phi^\dagger]^2$ (see Figs.~\ref{fig:NN} and \ref{fig:ENN}), as well as statistics on the eigenvalues and, as by the former, about the SSV.

In the Monte Carlo simulation for the JPDF of the SSV, we start from a randomly chosen configuration for each coordinate. As the JPDF of the SSV is invariant under permutation of the eigenvalues, the dynamics of the Monte Carlo process can be well understood by looking at the plot of the resulting SSV density. For a potential such as in Fig.~\ref{fig:potential} one will find that the density is supported on two disjoint intervals \footnote{Theoretically, for finite $N$, the density is positive everywhere since \eqref{jpdx} is always non-zero for any configuration as long as all SSV are mutually distinct. But in the appearing gap, the density is exponentially suppressed, such that even for rather small $N$ the probability to find a SSV there becomes vanishing small.}(see Fig.~\ref{fig:SVdensity}). In order that the Monte Carlo process finds the true equilibrium in finite time, it is important to choose the step-size of a random move large enough that it can go across the gap between the two intervals in a single step. A large step size will result in a low acceptance rate. Consequently, a large number of such steps will be needed. 

While the simulation of the JPDF \eqref{partition2} requires low computational effort, the determinant in \eqref{jpdx} is highly ill-conditioned, unless $g$ is very large. This requires to calculate the determinant using arbitrary-precision arithmetic with up to 500 digits. 

Since in each Monte Carlo move, we propose an update to only one singular value at a time, and therefore we change one row and one column of the determinant in \eqref{jpdx}, we can calculate the determinant of the new configuration from the old one.  Specifically, when we update the $k$-th SSV $x_k$, the ratio of the determinants can be expressed by  
\begin{equation}
    \frac{\det(\tilde{A})}{\det(A)}=(1+w_k)^2 -(A^{-1})_{kk}\vec{w}^{\,\dagger} \vec{v},
\end{equation}
where $A$ and $\tilde{A}$ are the matrices with the entries $A_{ij}=e^{-Ng(x_i-x_j)^2}$ and $\tilde{A}_{ij}=e^{-Ng(\tilde{x}_i-\tilde{x}_j)^2}$ respectively, and $\tilde{x}_i=x_i$ for $i\neq k$, while $\tilde{x}_k$ is the position of the $k$-th SSV after the proposed random move. $\vec{v}$ is the column vector with components $v_i=\tilde{A}_{ki}-A_{ki}$ and $\vec{w}=A^{-1}\vec{v}$ with components $w_i$. In this way, the change of energy of the new configuration can be calculated with complexity $\mathcal{O}(N^2)$. But for accepted updates, we have to calculate the inverse of the matrix $A$ with arbitrary-precision floating point numbers, which is time consuming.

For the full matrix Monte Carlo simulations we generate the JPDF given in \eqref{prob1} by sequentially proposing a random move to each matrix element $\phi_{ij}$. Since we change only one element at the time, we can calculate the difference in energy with respect to the old configuration in $\mathcal{O}(N)$ time when we keep track of the helper matrices $\psi=\phi^\dagger \phi$ and $\chi=\psi \phi^\dagger$, while an accepted move requires an update of all the elements in $\chi$ and has complexity $\mathcal{O}(N^2)$. 

As is often the case in Monte Carlo simulations, high probability configurations of $\phi$ in the $2N^2$ dimensional manifold are not intuitively obvious. During the simulation, we noticed that the ‘local’ moves detailed above are insufficient to prevent the simulation from becoming stuck in metastable
states. Physically motivated ‘global moves’ are needed to traverse phase space efficiently and
reduce the autocorrelation time \cite{NewmanBarkema1999,LandauBinder2021,SwendsenWang1987,Evertz1993,scalettar1991ergodicity,cohenstead2022fast}. We calculated the SSV from $\phi$ and compare the result with the direct SSV simulation. In both cases, we noticed that the density of SSV is supported on two disconnected intervals. However, while the number of SSV in one of the intervals is fluctuating in the direct SSV simulation from sample to sample, in the full matrix simulation it is fixed, depending on the initial matrix configuration. We assume that the full matrix simulation reaches equilibrium, subjected to the constraint of having a fixed certain number of SSVs in each of the intervals. We use the direct SSV simulation to generate an initial configuration for the full matrix simulation as follows: From an equilibrium sample of SSVs we produce a diagonal matrix $\Lambda$ and multiply it from the left and from the right with independently random unitary matrices to form the non-hermitian matrix $\phi$ (see \eqref{SVD}), and use this $\phi$ as the initial configuration for a full-matrix MC process. We then run a full matrix Monte Carlo process for each such initial configuration $\phi$, and average their equilibrium results with a weight depending on how often their initial SSV configuration appeared in the direct SSV simulation. By calculating the density of SSVs from the ensemble of full matrices thus created, we could exactly reproduce the same density we obtained from the direct SSV simulation, up to statistical fluctuations. It is reasonable to assume, that the same procedure also produces any other quantity we measure, like the density of eigenvalues. 

As a simple diagnostic of our MC code we verified that it reproduced known finite-$N$ exact results for the expectation values of some observables in Ginibre's ensemble
\begin{equation}\label{Ginibre}
 \tilde P_{\rm Ginibre}(\phi,\phi^\dagger)  = \left(\frac{Nm^2}{\pi}\right)^{N^2}e^{-Nm^2{\rm Tr}(\phi^\dagger\phi)}\,.
\end{equation}
For this simple Gaussian model one readily finds $\langle \frac{1}{N}{\rm Tr}(\phi^\dagger\phi)\rangle  = \frac{1}{m^2}$, $\langle \frac{1}{N}{\rm Tr}(\phi^\dagger\phi)^2\rangle = \frac{2}{m^4}$  and also the quantity
\begin{equation}\label{eq:comsq}
\left\langle \frac{1}{N}{\rm Tr}[\phi,\phi^\dagger]^2\right\rangle = \frac{2}{m^4}\left(1-\frac{1}{N^2}\right)
\end{equation}
which measures the deviation of $\phi$ from normality in Ginibre's ensemble. 
In particular, note that the last expectation value vanishes for $N=1$, as it should. Our code produced correctly all these results, including the finite-$N$ correction in the last expectation value \eqref{eq:comsq}.

\section{Numerical Results}\label{sec:results}

From Monte Carlo simulations described above, 
the violation of the Single Ring Theorem with sufficiently large $g$ is indisputable.
See, for example, Fig.~\ref{fig:qualitative}.
However, locating the precise critical value $g_{\rm crit}$ at which the distribution
becomes disjoint in the
`thermodynamic limit' $N\rightarrow \infty$ is a bit subtle
given the computational limits on accessible matrix dimensions--
simulations of adequate length for $N=64$ take roughly two weeks per parameter set
on a single core of a modern workstation.

\begin{figure}[t!]
\includegraphics[height=3.2in,width=3.20in,angle=-90]{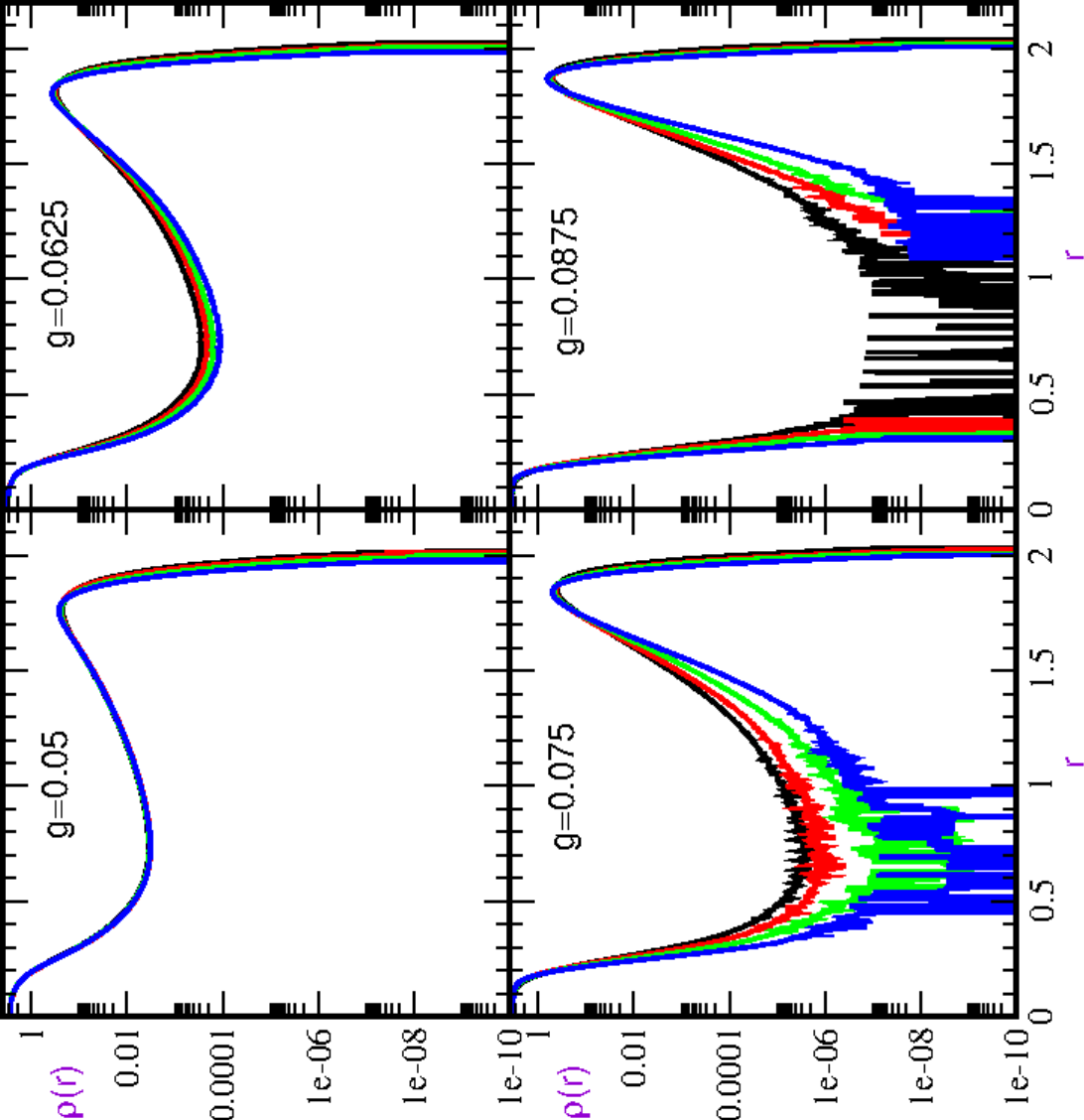}
\caption{
In this and in all subsequent figures in Section~\ref{sec:results} we display numerical results for the cubic potential in Fig.~\ref{fig:potential}. Here we show the eigenvalue distribution  $\rho(r)$ for different $g$, the coefficient of the commutator term. As $g$ increases, the distribution evolves into two disjoint regions of support. Values are shown for four matrix dimensions $N= 32$ (black), 40 (red), 48 (green), and 64 (blue). For $g \lesssim 0.0625$ finite size effects are minimal -- $\rho(r)$ is independent of $N$. However, for $g \gtrsim 0.0625$ the density in the intermediate region begins to fall as $N$ grows, a trend which becomes increasingly pronounced with larger $g$, suggesting $\rho(r) \rightarrow 0$ as $N \rightarrow \infty$. 
}
\label{fig:Pofr}
\end{figure}

\begin{figure}[t!]
\includegraphics[height=3.3in,width=3.30in,angle=0]{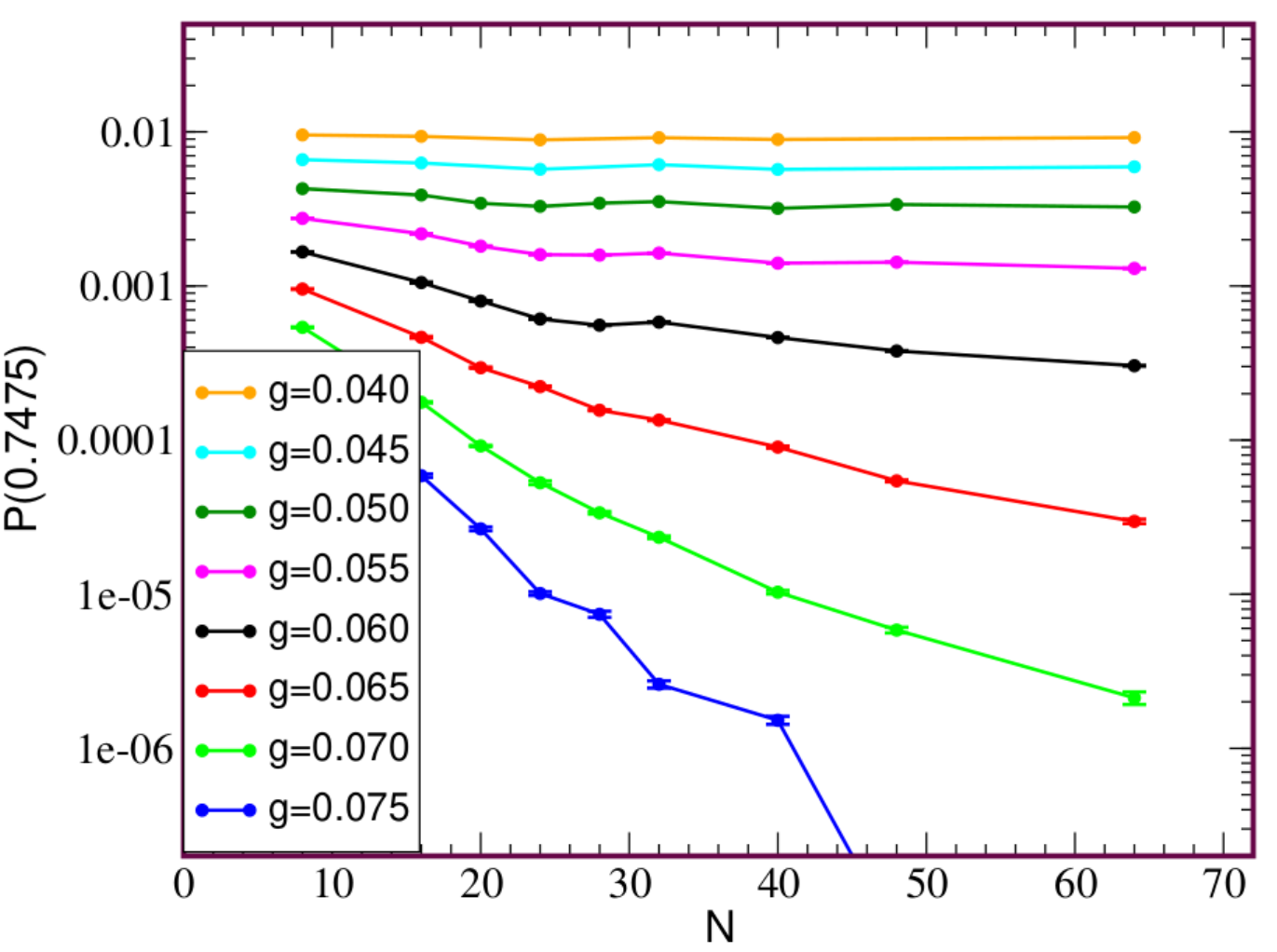}
\caption{
Scaling of the eigenvalue density $\rho(r)$ at the middle of the disk-annulus gap
$r \sim 0.7475$.
}
\label{fig:Plamlog}
\end{figure}

We will therefore provide several different `finite size scaling' approaches to 
extrapolation to $N=\infty$ and use those analyses to build up an 
estimate for $g_{\rm crit}$.

The ensemble \eqref{prob1} is rotation-invariant. Consequently, the average density $\rho(r)$ of complex eigenvalues $z_\alpha = x_\alpha + iy_\alpha$ of $\phi$ depends only $r=|z|=|x+iy|$. (In practice, we measure the number of eigenvalues in a very narrow ring of average radius r, and divide it by the area of the ring and by the total number $N$ of eigenvalues. This in effect also makes an angular average, and produces $\rho(r)$.) We begin by exhibiting $\rho(r)$ in Fig.~\ref{fig:Pofr} for four values of $g$, each for
four matrix dimensions $N=32, 40, 48, 64$.  For $g=0.0500$ there is no break
in $\rho(r)$, nor any visible change in the density as $N$ increases.
For $g=0.0625$ the density in the region between the two maxima is two orders of magnitude
smaller than for $g=0.0500$, and, moreover, now has a clear
tendency to shrink as $N$ grows.  Finally, for $g=0.0750$ and $g=0.0875$,
$\rho(r) \lesssim 10^{-5}$ and decreases significantly with growing
matrix size $N$.  Noting that the use of a logarithmic 
scale somewhat obscures changes in $\rho(r)$, these data suggest $g_{\rm crit} \lesssim 0.0625$.

 Finite-$N$ corrections typically swell the support of the average density of eigenvalues of random matrices by a small amount, which scales with some negative power of $N$. In particular, in models where the density is supported on a two dimensional area in the complex plane, at a point $z_0$ on the (infinite-$N$) boundary, typical finite-$N$ corrections render the density scaling as $\rho(z_0+\xi)\sim \mathrm{erfc}(\sqrt{2N}\xi)$, where $\xi$ denotes the distance to the boundary in the outwards normal direction. This universal behavior has been observed for non-Hermitian matrices \cite{rider2003limit,bender2010edge,cipolloni2021edge,campbell2025spectral,akemann2025spectral,liu2026repeated} as well as in the canonical normal matrix model \cite{AmeurHedenmalmMakarov2015,LR2016,AKM2019,HW2022,cronvall2025direct,charlier2025smallest}. In our case, we are looking for the critical $g$ at which a gap is formed. However, due to the finite-$N$ corrections at the two proximate boundaries, we are able to observe the gap in the eigenvalue density only when it is wide enough, that is, only when $g>g_\text{crit}$. Thus, rather than searching for the first formation of the gap directly, as in Fig.~\ref{fig:Pofr}, a better strategy for pinpointing the critical $g_{\rm crit}$ is to fix some radius $r_0$ from the origin, inside the region where the gap would form in the large-$N$ limit, fix a value of $g$, and measure the $N$ dependence of the density $\rho(r_0)$. If we find that $\rho(r_0)$ tends to a positive constant value as $N$ increases, that means $r_0$ is inside the large-$N$ limiting support of the density, that no gap was formed, and therefore $g<g_{\rm crit}$. If, on the other hand, we find that $\rho(r_0)$ decays exponentially to zero as $N\rightarrow\infty$, that means $r_0$ is in the gap and therefore $g> g_{\rm crit}$. We follow this strategy in Fig.~\ref{fig:Plamlog}, where we plot the eigenvalue density $\rho(r_0)$ as a function of matrix size $N$ at the fixed $r_0=0.7475$, chosen to be near the center of the disk-annulus gap, which would develop with $g$ (see Fig.~\ref{fig:qualitative}). Data are shown for eight values of $g$.  
At the four smallest $g=0.040, 0.045, 0.050, 0.055$, $\rho(r_0)$ appears nearly constant. However, at $g=0.060$ a clear (exponential) dependence develops.
Thus, according to the results plotted in Fig.~\ref{fig:Plamlog}, we estimate $0.055 \lesssim g_{\rm crit} \lesssim 0.060$, which is consistent with the estimate $g_{\rm crit} \lesssim 0.0625$ from the analysis of Fig.~\ref{fig:Pofr}.

\begin{figure}[t!]
\hspace*{-0.10in}
\includegraphics[height=3.2in,width=2.8in,angle=-90]{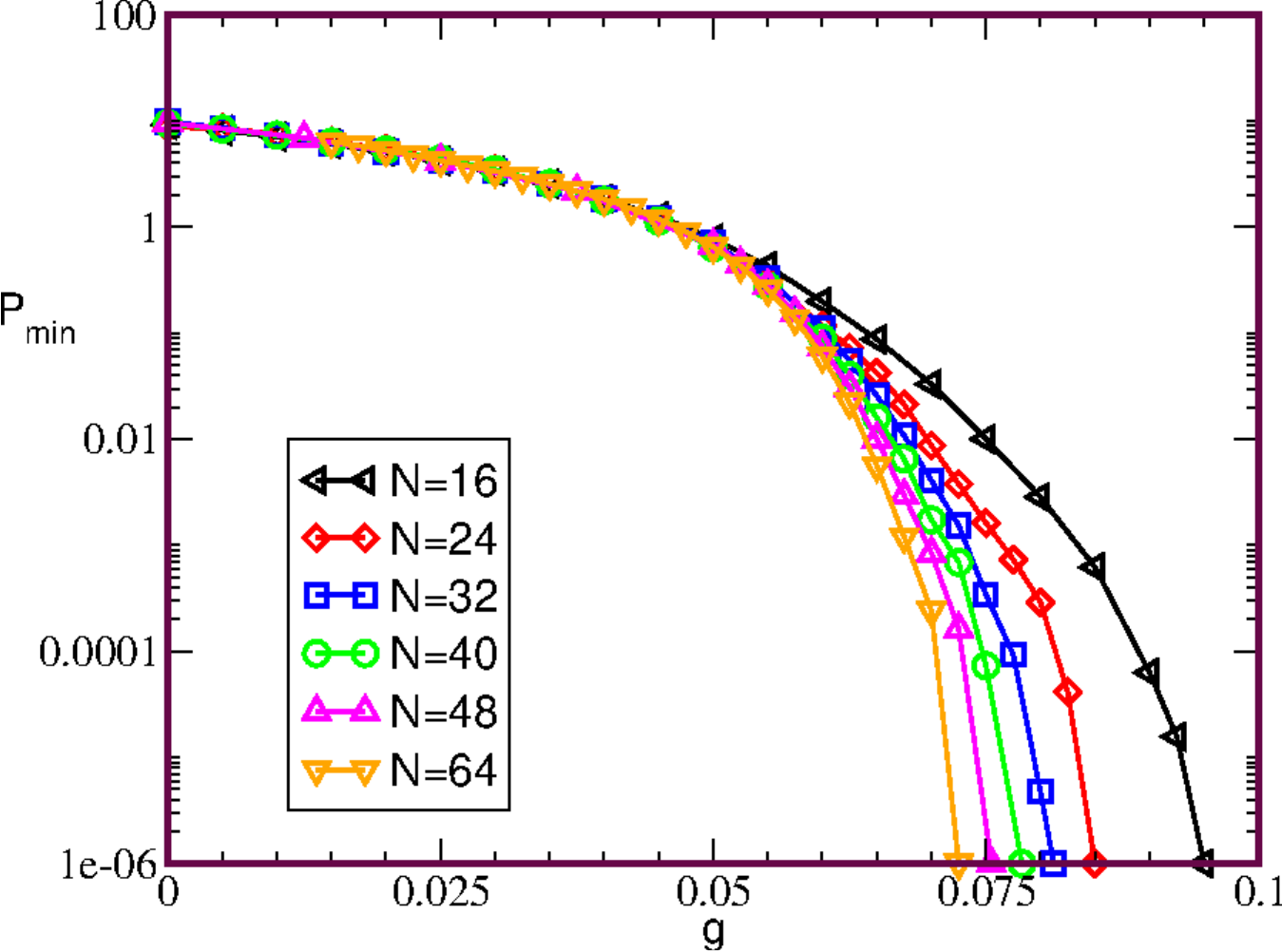}
\caption{
Plot of the minimum value $P_{\rm min}$ of the eigenvalue density $P(|z|)\equiv \rho(r)$ in the gap between the disk and annulus for six matrix sizes $N=16, 24, 32, 40, 48, 64$. A value $P_{\rm min}=10^{-6}$ is displayed at the first $g$ value for which the minimum value is zero (an eigenvalue bin is empty). 
}
\label{fig:plammin}
\end{figure}

In Fig.~\ref{fig:plammin} we show the smallest measured value $\rho_{\rm min}$ of the
eigenvalue distribution
$\rho(r)$ as a function of $g$.
$\rho(r)$ diminishes roughly linearly with $g$ for small $g$, and then
turns over more sharply.
These data suggest that $g_{\rm crit} \lesssim 0.0750$, in agreement with the
trends in Fig.~\ref{fig:Plamlog}.

\begin{figure}[t!]
\includegraphics[width=3.0in,height=3.0in,angle=-90]{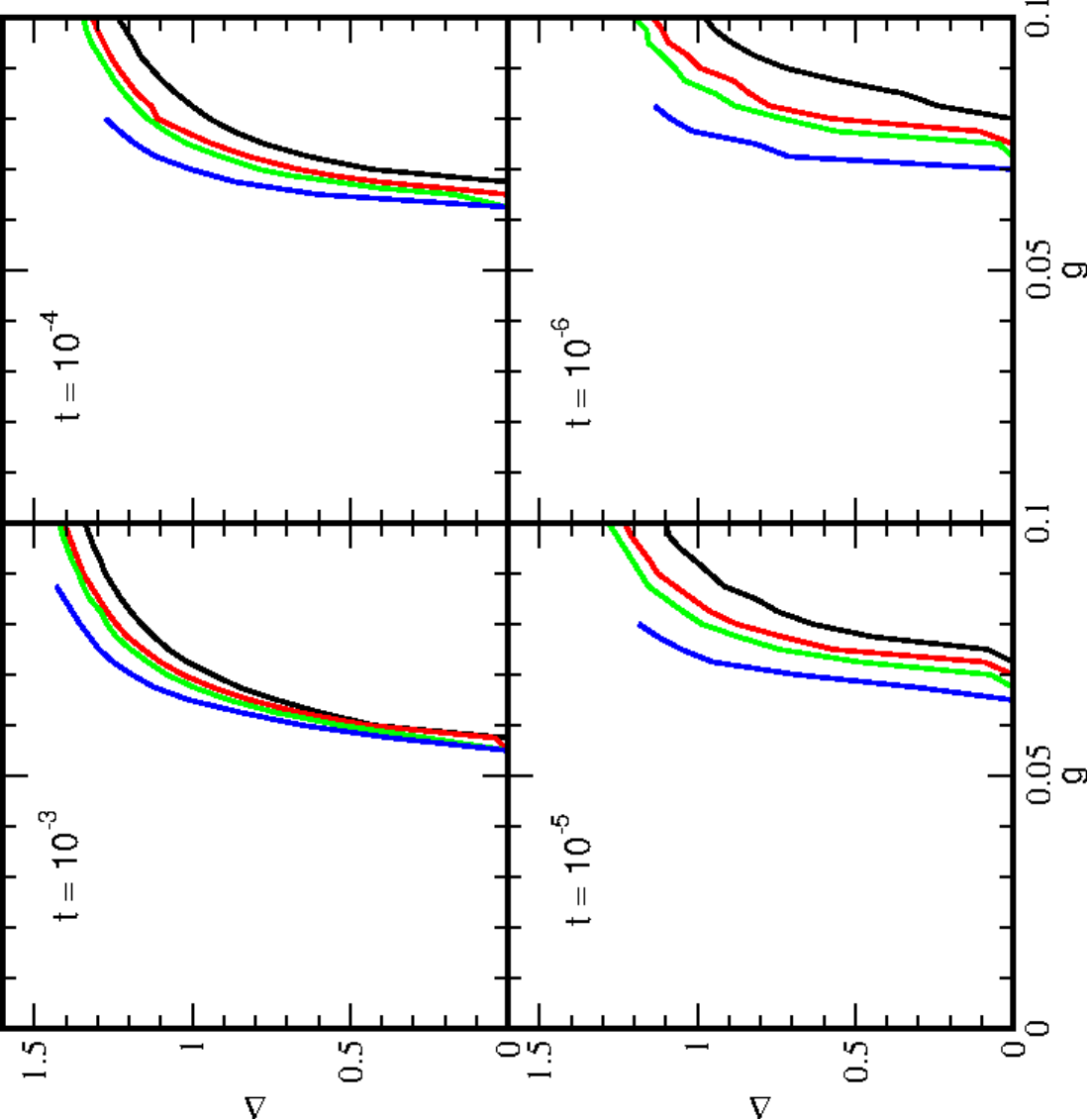}
\caption{
Plot of the size of the gap $\Delta$ of the eigenvalue density in the region between disk and annulus.  The gap is defined by counting the number of bins whose density of states is less than some threshold $t$ and multiplying by the bin width.   In each panel, data are shown for four matrix dimensions, $N=24$ (black), $32$ (red), $40$ (green), $64$ (blue).  These appear right to left in each panel. See text.
The critical value of the commutator term $g_{\rm crit}$ inferred from the largest matrix dimension, $N=64$, and smallest threshold, $t=10^{-6}$, is consistent with those of Figs.~\ref{fig:Plamlog} and \ref{fig:plammin}.
}
\label{fig:gap}
\end{figure}

We can also track the size of the `gap' $\Delta$ between the nonzero densities in the
central disk and the outer annulus by defining a threshold $t$ and counting the
interval of $r$ for which $\rho(r) < t$.
Data for $\Delta$ are shown in 
Fig.~\ref{fig:gap} for $N=24, 32, 40, 64$.
One can interpret the value of $g$ for which $\Delta$ becomes nonzero as
the critical point, although this analysis relies on the {\it ad hoc} choice
of threshold $t$. 
$\Delta$ increases as $N$ increases- eigenvalues intruding into the region between central disk and outer annulus are finite size artifacts which reduce $\Delta$.  Likewise, the onset of the gap shifts to higher values of $g$ as the choice of $t$ is reduced. This analysis is subject not only to finite size effects, but also involves the {\it ad hoc} choice of threshold $t$.  Hence one can only infer a fairly broad range,
$0.05 \lesssim g_{\rm crit} \lesssim 0.07$ for the critical coupling.

\begin{figure}[t!]
\includegraphics[width=3.5in,height=3.0in]{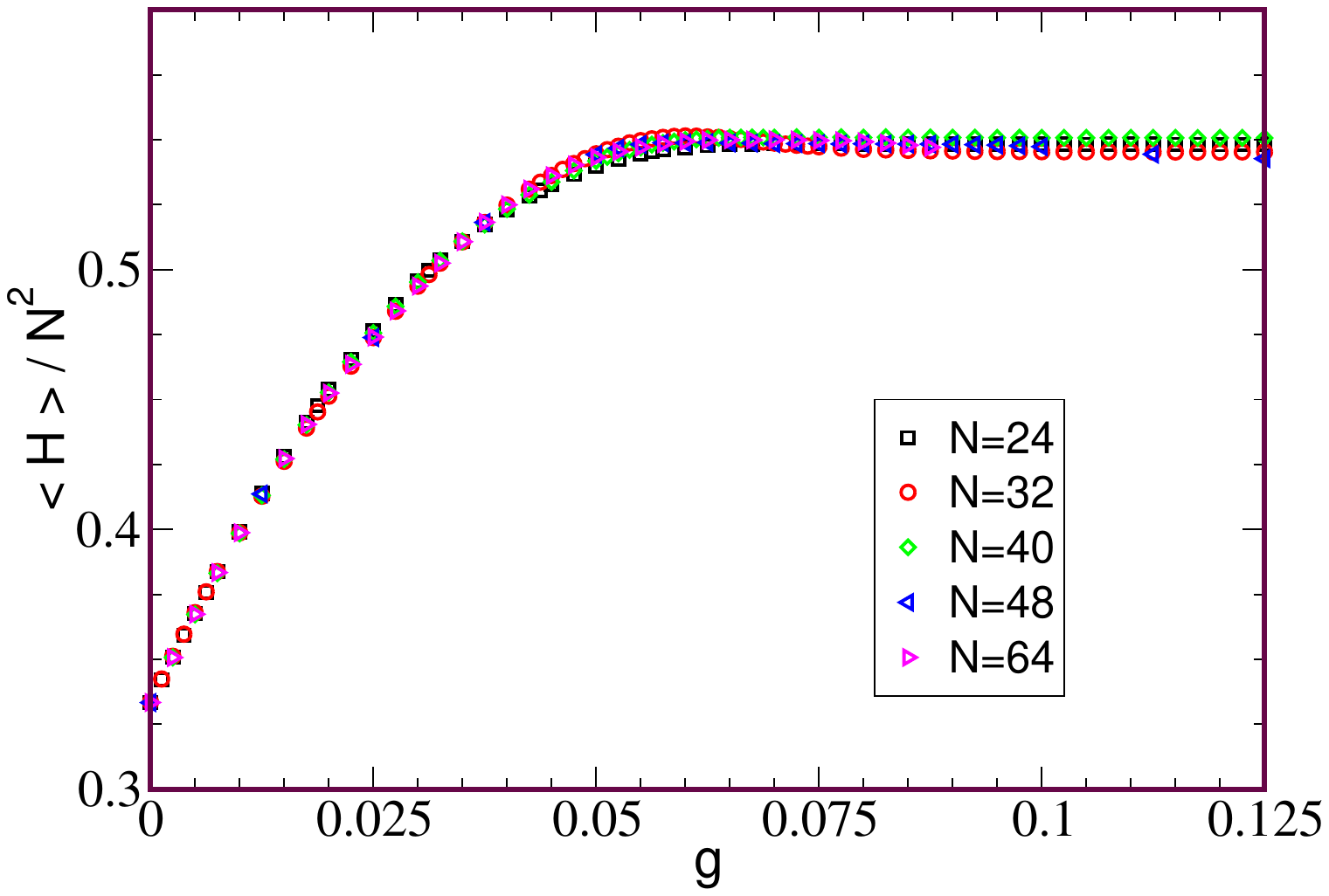}
\caption{
Average energy $\langle H \rangle = \langle N{\rm Tr} V + Ng{\rm Tr}[\phi,\phi^\dagger]^2\rangle$ as a function of $g$. The energy is roughly independent of $g$ for $g \gtrsim 0.055$, that is, in the region where the eigenvalue distribution occupies two disjoint regions.
}
\label{fig:E}
\end{figure}

\begin{figure}[t!]
\includegraphics[width=3.5in,height=3.0in]{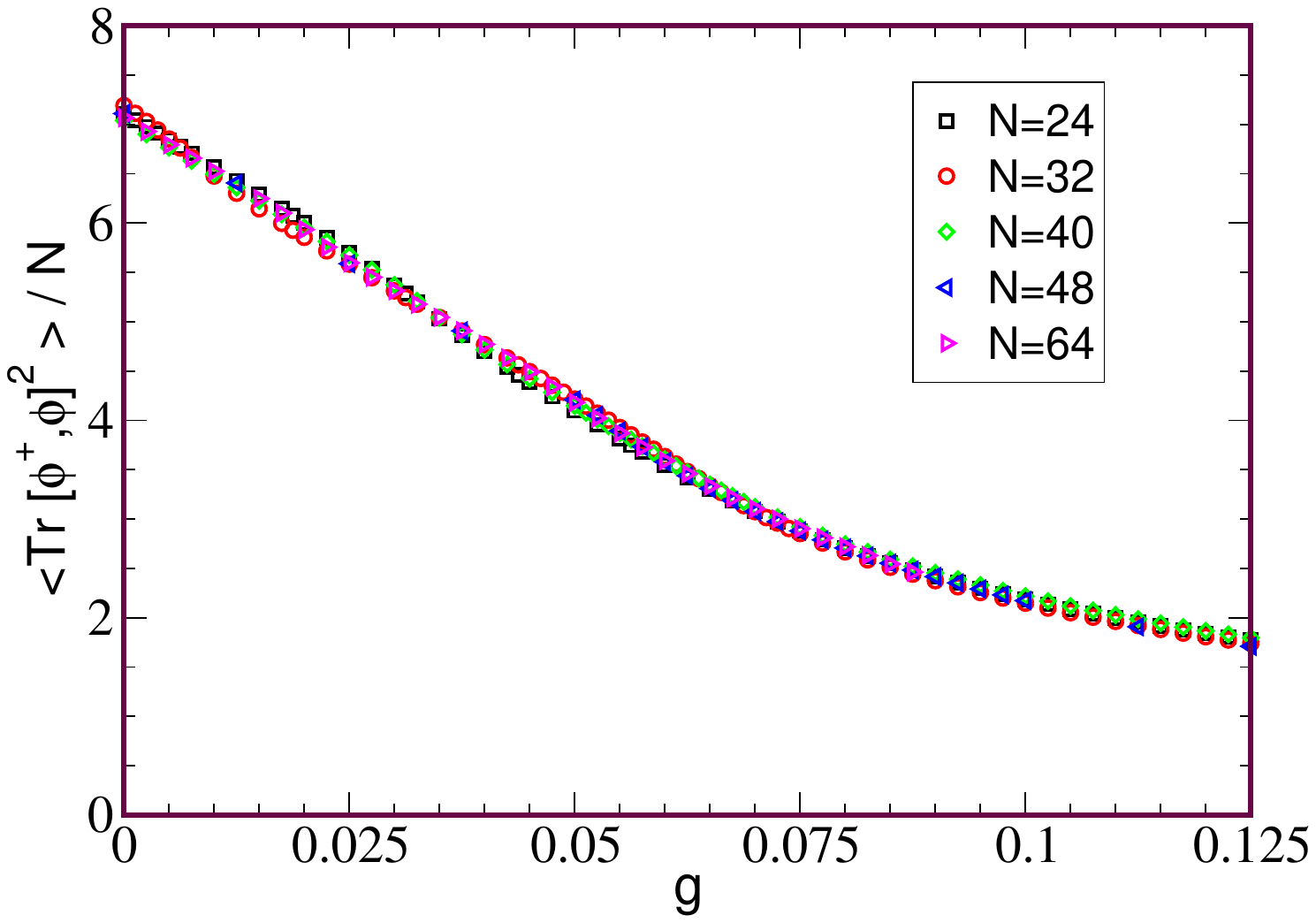}
\caption{
Non-normality (trace of the commutator square) versus $g$.  
An initial linear decrease at small $g$ seems to curve to smaller slope at 
large $g$.
}
\label{fig:NN}
\end{figure}

\begin{figure}[t!]
\includegraphics[width=3.5in,height=3.0in]{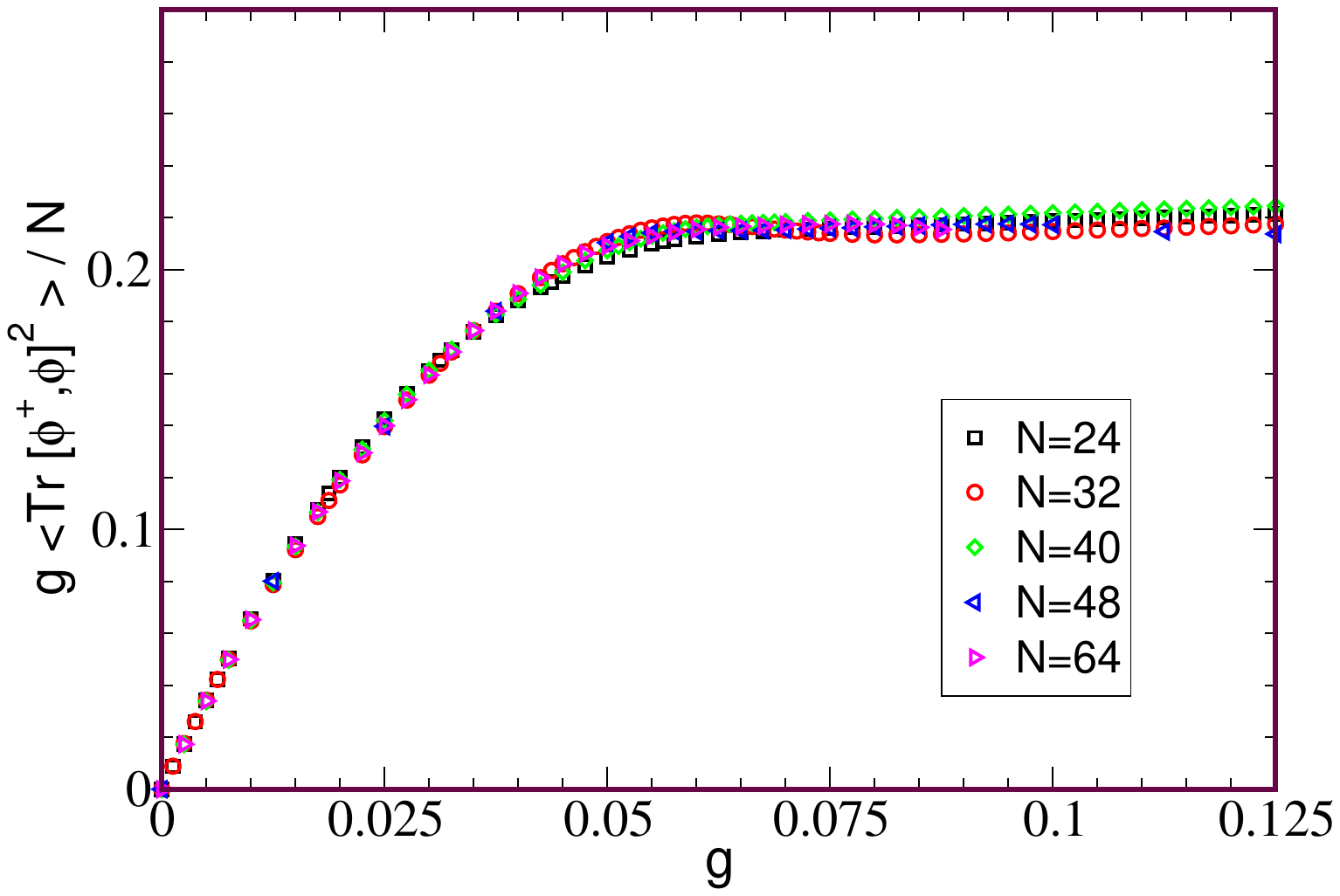}
\caption{
Non-normality energy versus $g$.  
This is the data of Fig.~\ref{fig:NN} multiplied by $g$.
As for the total
energy shown in Fig.~\ref{fig:E}, this component 
is roughly independent of $g$ for $g \gtrsim 0.055$, that is, in the
region where the eigenvalue distribution occupies two disjoint regions.
}
\label{fig:ENN}
\end{figure}

Typically in Monte Carlo simulations of statistical mechanical systems the energy is rather smooth
and not a precise signal of critical points.  Instead, various
derivatives (the `specific heat' in statistical physics) are used to pull out
incipient singularities.  Nevertheless, the energy often
shows some sort of cross-over in the vicinity of transitions.
We see precisely that phenomenon in an 
analysis of the $g$ dependence 
of different components of the effective matrix potential (the exponent of \eqref{prob1}) in Figs.~\ref{fig:E}, \ref{fig:NN}, and \ref{fig:ENN}.
While there is a clear change in slope in these quantities as $g$ increases, the relative smoothness precludes a precise determination of $g_{\rm crit}$.  Clearly, however, these data are consistent with our prior estimates.

In summary, we believe the measuring strategy of $g_{\rm crit}$ leading to Fig.~\ref{fig:Plamlog} is the most sensitive and reliable, putting our result at $g_{\rm crit} = 0.055\pm 0.005$. 

Fragility of the SRT under the perturbation ${\rm Tr}[\phi,\phi^\dagger]^2$, that is, the relative smallness of $g_{\rm crit}$ compared to $1$, can be justified as follows. It is natural to expect the SRT-breaking transition to occur when the `penalty term' $\langle Ng{\rm Tr}[\phi,\phi^\dagger]^2\rangle $ in \eqref{partition1} becomes comparable with the `potential energy' $\langle N {\rm Tr} V(\phi^\dagger\phi)\rangle$. 
Let's concentrate on $g=0$. In the unperturbed ensemble \eqref{prob} the matrices $\phi$ are almost surely non-normal. That is, ${\rm Tr}[\phi,\phi^\dagger]^2 >0$ almost surely for any $\phi$ drawn from \eqref{prob} (recall that $[\phi,\phi^\dagger]$ is a hermitian matrix). Thus, upon averaging, $\left\langle \frac{1}{N}{\rm Tr}[\phi,\phi^\dagger]^2\right\rangle >0$. 

Notice further that all curves shown in Figs.~\ref{fig:E} and \ref{fig:NN} fall approximately on top of each other, exhibiting large-$N$ scaling already for the moderate matrix sizes shown.

Thus, in the case of the cubic potential shown in Fig.~\ref{fig:potential}, we see from Fig.~\ref{fig:NN} that $\left\langle N{\rm Tr}[\phi,\phi^\dagger]^2\right\rangle _{g=0} \approx 7 N^2$. Similarly, we see from Fig.~\ref{fig:E} that in this case the total `energy' $\langle H\rangle = \langle N {\rm Tr} V\rangle_{g=0}\approx 0.33 N^2$, a factor of about $20$ {\em smaller} than $\left\langle N{\rm Tr}[\phi,\phi^\dagger]^2\right\rangle _{g=0}$, which by continuity tells us that $g_{\rm crit}$ should be around $1/20 = 0.05$. Around the actual $g_{\rm crit}\approx 0.055$ we have $\langle H \rangle \approx 0.55 N^2$, $\left\langle Ng{\rm Tr}[\phi,\phi^\dagger]^2\right\rangle \approx 0.22N^2$ and therefore $\langle V \rangle = \langle H \rangle - \langle Ng{\rm Tr}[\phi,\phi^\dagger]^2\rangle\approx 0.33 N^2$, rendering the breaking term $\langle Ng{\rm Tr}[\phi,\phi^\dagger]^2\rangle$ comparable with the potential term $\langle V \rangle$, in accordance with our gross prediction.

We expect our gross estimate $g_{\rm crit}\approx 0.05$ to be valid also in the large-$N$ limit because, as was mentioned above, the data in Figs.~\ref{fig:E} - \ref{fig:ENN} appear to exhibit large-$N$ scaling already for the moderate matrix sizes we simulated.

In addition to the location of critical points, it is often of interest
to characterize more precisely the nature of the singularity-  is the
transition first or second order?  When sufficiently large systems are accessible, finite
size scaling can address this issue\cite{Challa1986,LeeKosterlitz1991}. For example in a magnetic phase transition the
susceptibility (a measure of fluctuations in the order parameter) will scale with the volume $\chi \propto V=L^{d}$ at a first order transition
but with non-trivial critical exponents $\chi \propto L^{\gamma/\nu}$ at a second
order transition.  More sophisticated observables like the Binder cumulant, an appropriate
dimensionless ratio of powers of the order parameter, 
can exhibit a sign change if the transition is first order.  Finally, distributions of 
the order parameter are qualitatively different-  bimodal at a first order phase change, and
a single peak at a second order phase change.
In the discussion above we used several methods to locate the critical point, including the
eigenvalue density in the gap and the size of the gap, as measured by the extent
of the region where the density falls below a threshold value.  However, neither of these
measures constitutes an order parameter in the traditional sense.  This precludes us
from adopting the known approaches to determining the nature of the singularity.
Thus the question of the order of the transition with increasing $g$ remains open.

\FloatBarrier
\section{Conclusion}\label{sec:conclusion}
In this paper we have verified numerically the qualitative argument originally given 
in \cite{feinberg1997non, feinberg2001single} and reviewed in Section \ref{Section:Introduction}, that the Single Ring Theorem would break once the unitary matrices $U$ and $V$ in \eqref{SVD} become correlated. In the concrete model \eqref{prob1} with the potential shown in Fig.~\ref{fig:potential}, SRT breaking occurs already at that relatively small value of $g_{\rm crit}$ around $0.055$. Thus, the SRT is rather fragile with respect to the perturbation $g{\rm Tr}[\phi,\phi^\dagger]^2$. 

We have also studied the gas of singular values \eqref{jpdx} associated with \eqref{prob1}. It is a non-interacting Fermi gas, confined to the positive half line.
An important feature of this gas is that its interparticle spacing statistics changes as the flow parameter $g$ changes from Wigner-Dyson repulsion at the starting point $g=0$ of the flow, to non-repulsive Poisson statistics at the $g\rightarrow\infty$ end point of the flow, where the ensemble \eqref{prob1} concentrates on normal matrices. 

However, this continuous change in interparticle statistics, which is interesting in its own right, has nothing to do with the critical point along the flow at which the SRT breaks, which occurs (for the concrete cubic potential mentioned above) at a small critical value of $g$. Indeed, as one sweeps the parameter $g$ through the SRT breaking point, the density of SVs shown in Fig. \ref{fig:SVdensity} barely changes. This rigidity of the density of SVs through the SRT breaking transition is in accordance with the argument of \cite{feinberg1997non, feinberg2001single}, which assumed a fixed spectrum of SVs, and only increased correlation between $V$ and $U$, rendering $W=VU$ more and more diagonal. In addition to the rigidity of the density of SVs through the SRT breaking transition, also the empirical unfolded spacing distribution $P(s)$ shown in Fig. \ref{fig:SVspdistN2x} barely changes. 

Moreover, this change in statistics of interparticle spacings is a universal phenomenon, occurring for any potential $V(x)$ in \eqref{jpdx}, including matrix potentials $V(\phi^\dagger\phi)$ which produce only a single disk, such a Ginibre's potential $V(x) = x$ or a potential like $V(x) = x + x^2$ which does not have a minimum on the positive half line, except at $x=0$.

\appendix
\renewcommand{\thesection}{\Alph{section}}
\renewcommand{\thesubsection}{\thesection.\arabic{subsection}}

\section{Normal Matrices and their Singular Values}\label{AppA}

An $N\times N$ normal matrix $\phi$ commutes with its hermitian adjoint $[\phi,\phi^\dagger]=0$, from which it follows that it is diagonalizable by a unitary matrix $U$:
\begin{equation}\label{normal}
    \phi = U^\dagger Z U\,,\quad Z={\rm diag}(z_1,z_2,\ldots, z_n),
\end{equation}
where $z_i$ are the $N$ complex eigenvalues of $\phi$. Thus, 
\begin{equation}
 \phi^\dagger\phi = U^\dagger~ {\rm diag} (|z_1|^2,|z_2|^2,\ldots, |z_N|^2)~ U
\end{equation}
meaning that the singular values of $\phi$, 
\begin{equation}\label{SVD-normal}
    \Lambda = {\rm diag} (\lambda_1,\lambda_2,\ldots, \lambda_N) = {\rm diag} (|z_1|,|z_2|,\ldots, |z_N|)
\end{equation}
are simply the moduli of its corresponding eigenvalues, as was discussed in the Introduction. 

In this respect, normal matrices can be thought of simply as analytical extensions of hermitian matrices, whose real eigenvalues are allowed to wander off into the complex plane. Consequently, the Cartesian integration measure over \emph{unitary-invariant} normal matrix ensembles is factorized as
\begin{equation}\label{measure}
d\phi d\phi^\dagger = d\mu(U)~ |\Delta (Z)|^2~ d^2z_1\ldots d^2z_N
\end{equation}
where $d\mu(U)$ is the Haar measure over the coset $U(N)/U(1)^N$, and 
\begin{equation}\label{vandermonde}
\Delta (Z)  = \prod_{i>j} (z_i-z_j) = \det_{ij} z_i^{j-1}   
\end{equation}
is the Vandermonde determinant of complex eigenvalues. Thus, unitary rotations are statistically independent of eigenvalues, in complete analogy with random unitary-invariant hermitian matrix ensembles. 

Consider unitary-invariant normal matrix models which are also rotational invariant, namely, $\phi$ and $e^{i\alpha}\phi$ are equally probable for any real $\alpha$. The ``canonical" probability distribution for such ensembles is of the general form \cite{Chau}
\begin{equation}\label{rotational}
P(\phi,\phi^\dagger)  \propto e^{-N {\rm Tr} V(\phi^\dagger\phi)}
\end{equation}
where $V$ is a potential function. The joint probability distribution function (JPDF) of eigenvalues of such matrices is therefore 
\begin{eqnarray}\label{zpdf}
&&P(z_1,\ldots, z_N)\, d^2z_1\ldots d^2z_N =\nonumber{}\\ &&\frac{1}{\mathcal{Z}_N} |\Delta (Z)|^2~ e^{-N\sum_i V(|z_i|^2)} d^2z_1\ldots d^2z_N
\end{eqnarray}
where $\mathcal{Z}_N$ is the normalizing partition function.

Eigenvalues can be expressed in polar form as $z_i=\lambda_i e^{i\theta_i}$, and correspondingly $d^2z_i = \lambda_i d\lambda_i d\theta_i$, where we used \eqref{normal} and \eqref{SVD-normal}, and where the polar angle $\theta_i$ is distributed uniformly over $[0,2\pi)$. Thus, we obtain from \eqref{zpdf} the JPDF of singular values as 
\begin{eqnarray}\label{svpdf}
   &&Q(\lambda_1,\lambda_2,\ldots, \lambda_N)\,  = \frac{1}{\mathcal{Z}_N}\,e^{-N\sum_i V(\lambda_i^2)} \lambda_1 \cdots \lambda_N\nonumber{}\\&&  \int_0^{2\pi}\frac{d\theta_1}{2\pi}\cdots \frac{d\theta_N}{2\pi} |\Delta (\lambda_k e^{i\theta_k})|^2\,.
\end{eqnarray}
In order to average $|\Delta|^2$ over the polar angles, we expand each determinant in permutations 
\begin{equation}\label{permutation}
\Delta(Z) = \det_{ij} (\lambda_i e^{i\theta_i})^{j-1} = \sum_{\sigma\in S_N} (-1)^\sigma \prod_{k=1}^N (\lambda_k e^{i\theta_k})^{\sigma(k)-1}    
\end{equation}
and substitute into \eqref{svpdf}, thus obtaining a double sum over permutations. Only the diagonal terms of this double sum survive averaging over the angles, and we end up with the permanent
\begin{equation}\label{permanent}
    \langle |\Delta(Z)|^2\rangle_{\theta_1,\ldots,\theta_N} = \sum_{\sigma\in S_N} \prod_{k=1}^N(\lambda_k^2)^{\sigma(k)-1} = {\rm Perm}_{k,l}(\lambda_l^2)^{k-1}
\end{equation}
of the matrix $(\lambda_l^2)^{k-1}$. Thus, finally, 
\begin{eqnarray}\label{svpdf-final}
    &&Q(\lambda_1,\lambda_2,\ldots, \lambda_N)\, d\lambda_1\cdots  d\lambda_N = \nonumber {}\\
    &&\frac{1}{2^N{\mathcal{Z}_N}}\,e^{-N\sum_i V(\lambda_i^2)}\,\left( {\rm Perm}_{k,l}(\lambda_l^2)^{k-1}\right) d\lambda_1^2\cdots  d\lambda_N^2\,.\nonumber{}\\
\end{eqnarray}
An important feature of \eqref{svpdf-final} is that in contrast to the quadratic eigenvalue repulsion in \eqref{zpdf}, there is no repulsion whatsoever among singular values. This is so because eigenvalues can share the same radius $\lambda$, but repel each other along the circle. This feature is manifested to its extreme in ensembles of unitary matrices, such as the circular unitary ensemble (CUE), where all eigenvalues lie on the unit circle.

\section{Free Fermions on the Half-Line}\label{AppB}

In this Appendix we provide some technical details required for establishing the fact that the singular values of $\phi$ in Section \ref{sec:Fermi} form a gas of non-interacting fermions living on the positive half-line. In order to keep the discussion in this Appendix as simple as possible, we shall set in this Appendix the confining potential $V$ in \eqref{jpdx} to zero. 

In the absence of a confining potential, we should imagine the system confined in a big box or compactified on a large ring, in order to prevent the particles flying off to infinity, keeping \eqref{jpdx} normalizable.

\subsection{A Single Particle}

Consistent, probability conserving formulation of the quantum mechanics of a particle living on the positive half-line requires that the probability current density vanishes at the origin\footnote{More exotic self-adjoint extensions leading to conserved non-vanishing persistent probability current are irrelevant for our discussion and are therefore excluded.}. The simplest boundary conditions, among the one-parameter family of boundary conditions imposed on the wave function at the origin which guarantee this requirement, are the null Dirichlet ($\psi_D(0) = 0)$ and Neumann ($\psi_N'(0) = 0)$ boundary conditions. Particles subjected to Dirichlet boundary conditions are repelled from the origin (the probability density at the origin vanishes in this case), whereas particles subjected to Neumann boundary conditions are reflected by the origin back in the positive direction. 

The basis of eigenstates of the free Dirichlet Hamiltonian $\hat{\cal H}_D=\hat p_D^2$ consists of the functions $D_p(x) = \sqrt{\frac{2}{\pi}} \sin (px)\,, (p>0)$. Similarly, the basis of eigenstates of the free Neumann Hamiltonian $\hat{\cal H}_N=\hat p_N^2$ consists of the functions $N_p(x) = \sqrt{\frac{2}{\pi}} \cos (px)\,, (p\geq 0)$.

The path integral representation in imaginary (Euclidean) time $\tau$ of the propagation amplitude $G_D(x,y;\tau)$ of Dirichlet particles between two points on the positive half line is the sum over trajectories of Brownian random walkers which avoid the origin altogether, while the corresponding amplitude $G_N(x,y;\tau)$ for Neumann particles 
is the sum over Brownian trajectories which may be reflected at the origin any positive number of times \cite{HL1} (see also \cite{HL2}). One can readily calculate $G_D(x,y;\tau)$ and $G_N(x,y;\tau)$ in terms of Fourier-sine and Fourier-cosine integrals, respectively, and obtain
\begin{eqnarray}\label{DN1}
    G_D(x,y;\tau) &=& \langle x|e^{-\tau \hat p_D^2} |y\rangle = \frac{1}{\sqrt {4\pi\tau}}\left(e^{-\frac{(x-y)^2}{4\tau}} - e^{-\frac{(x+y)^2}{4\tau}}\right)\nonumber\\{}\nonumber\\
    G_N(x,y;\tau) &=& \langle x|e^{-\tau \hat p_N^2} |y\rangle = \frac{1}{\sqrt {4\pi\tau}}\left(e^{-\frac{(x-y)^2}{4\tau}} + e^{-\frac{(x+y)^2}{4\tau}}\right)\,.\nonumber\\{}
\end{eqnarray}
It is straightforward to check that both amplitudes satisfy the heat equation subjected to the initial value $\delta(x-y)$. Furthermore, note that $G_D(0,y;\tau) = G_D(x,0;\tau) =0$, whereas $\partial_x G_N(x,y;\tau)_{x=0} = \partial_y G_N(x,y;\tau)_{y=0} =0$, as they should. 

The average of the two expressions in \eqref{DN1}
\begin{equation}\label{DN}
G_{DN}(x,y;\tau) = \langle x|\frac{1}{2}(e^{-\tau \hat p_D^2} + e^{-\tau \hat p_N^2} )|y\rangle = \frac{1}{\sqrt {4\pi\tau}} e^{-\frac{(x-y)^2}{4\tau}}
\end{equation}
is proportional to the Gaussian factor in the determinant in \eqref{jpdx}, implying that gas of singular values is an equally weighted mixture of Dirichlet and Neumann particles, diffusing freely (for $V=0$) along the positive axis. More precisely, \eqref{DN} gives the total propagation amplitude between the two points as the sum over all Brownian trajectories connecting these points, and restricted to the positive half-line, which may be reflected at the origin any non-negative number of times. 

$G_{DN}(x,y;\tau)$ should not be confused, of course, with the corresponding propagation amplitude 
\begin{equation}\label{whole}
G(x-y;\tau) = \langle x|e^{-\tau \hat p^2}|y\rangle = \frac{1}{\sqrt {4\pi\tau}} e^{-\frac{(x-y)^2}{4\tau}}
\end{equation}
of a free particle living on the {\em whole} real line, which is identical to it in form. 

In particular, the propagator $G(x-y;\tau)$ has the semi-group property $\int\limits_{-\infty}^\infty dy G(x-y;\tau_1)G(y-z;\tau_2) = G(x-z;\tau_1 + \tau_2)$ on the whole real line, as does each of the propagators in \eqref{DN1} on the positive half-line. However, their average \eqref{DN}, does not have this property on the positive half-line, because it is spoiled by the $G_NG_D$ cross-terms in $\int_0^\infty dy\, G_{DN}(x,y;\tau_1) G_{DN}(y;z;\tau_2)$. 

For technical reasons, it is convenient to make a direct sum of the two individual D and N Hilbert spaces and define the free propagator, or transfer matrix in this larger space to be 
\begin{equation}\label{transfer-doubled}
\hat {\cal T}^{(0)}(\tau) = \left ( \begin{array}{cc} e^{-\tau \hat p_D^2} 
     & 0 \\
    0 & e^{-\tau \hat p_N^2} 
\end{array}\right)
\end{equation}
(where the superscript $(0)$ indicates free particle), with matrix elements 
\begin{equation}\label{transfer-doubled-xy}
\langle x|\hat {\cal T}^{(0)}(\tau)| y\rangle = \left ( \begin{array}{cc}  G_D(x,y;\tau) 
     & 0 \\
    0 &  G_N(x,y;\tau)
\end{array}\right)
\end{equation}
which has, of course, the semi-group property $\int\limits_0^\infty dy \langle x|\hat {\cal T}^{(0)}(\tau_1)| y\rangle \langle y|\hat {\cal T}^{(0)}(\tau_2)| z\rangle  = \langle x|\hat {\cal T}^{(0)}(\tau_1+\tau_2)| z\rangle$ of a proper propagator on the positive axis. 
It would be further convenient to introduce the notation 
\begin{equation}\label{transfer-trace}
    \hat T^{(0)}(\tau) = \frac{1}{2}{\rm tr}\,\hat {\cal T}^{(0)}(\tau) =\frac{1}{2}(e^{-\tau \hat p_{D}^2} + e^{-\tau \hat p_{N}^2} )
\end{equation}
for the average of the Dirichlet and Neumann propagators, with matrix element 
\begin{equation}\label{transfer-trace-xy}
    \langle x|\hat T^{(0)}(\tau)| y\rangle = G_{DN}(x,y;\tau) = \frac{1}{\sqrt {4\pi\tau}} e^{-\frac{(x-y)^2}{4\tau}}\,.
\end{equation}
We shall refer to \eqref{transfer-trace} and \eqref{transfer-trace-xy} as the {\em quasi-propagator} of the mixed Dirichlet-Neumann particle and its matrix element. The qualifier ``quasi-" implying, as was explained above, that it is not a proper propagator on the positive half-line, despite its identical form to the free propagator on the whole real line. 

Of course, the quasi-propagator $G_{DN}(x,y;\tau)$ is obtained by restricting the spatial coordinates of the free propagator $G(x-y;\tau)$ to positive values. From the point of view of the large graded Hilbert space on which \eqref{transfer-doubled} is defined, the identitical form of these two objects can be traced back to the fact that the two momentum eigenstates $e^{\pm ipx}$ on the whole real line are proportional to the simple linear superposition $N_p(x)\pm iD_p(x)$ of the Dirichlet and Neumann eigenstates (upon extending the latter from the positive half-line to the whole line).

\subsection{Gas of Free Fermions on the Half-Line}

Let us now move to the $N$ particle case and show that \eqref{jpdx} describes a non-interacting Fermi gas occupying the positive half-line, consisting of an even mixture of Dirichlet and Neumann particles. To this end, we form from the $N$-particle position eigenstates
\begin{equation}\label{Nket}
|\Vec{x}\rangle = \prod_{k=1}^N |x_k\rangle = |x_1, x_2, \ldots, x_N\rangle  
\end{equation}
the complete basis of antisymmetric position eigenstates 
\begin{eqnarray}\label{Nketasym}
    |\Vec{x}\rangle_A &=& \frac{1}{\sqrt{N!}}\sum_{\sigma\in S_N} (-1)^\sigma \prod_{k=1}^N |x_{\sigma(k)}\rangle\nonumber\\ &=& \frac{1}{\sqrt{N!}}\sum_{\sigma\in S_N} (-1)^\sigma |x_{\sigma(1)},\ldots, x_{\sigma(N)}\rangle\,,
\end{eqnarray}
(where $(-1)^\sigma$ is the signature of the permutation $\sigma$), which fulfill the orthogonality relation
\begin{eqnarray}\label{Noverlap}
     {}_A\langle \Vec{x}|\Vec{y}\rangle_A
    &=&\sum_{\sigma\in S_N} (-1)^\sigma \prod_{k=1}^N \delta\left( x_k-y_{\sigma(k)}\right)\nonumber\\ &=& \det_{ij} \delta(x_i-y_j)\,.
\end{eqnarray}
    
Following \eqref{transfer-doubled} and \eqref{transfer-trace}, let us denote the propagator of the $k$-th particle as
\begin{equation}\label{transfer}
\hat {\cal T}_k^{(0)}(\tau) = \left ( \begin{array}{cc} e^{-\tau \hat p_{kD}^2} 
     & 0 \\
    0 & e^{-\tau \hat p_{kN}^2} 
\end{array}\right)
\end{equation}
and its contribution to the quasi-propagator of the Dirichlet-Neumann mixture by 
\begin{equation}\label{transfer-mixture}
    \hat T_k^{(0)} = \frac{1}{2}(e^{-\tau \hat p_{kD}^2} + e^{-\tau \hat p_{kN}^2} )\,.
\end{equation}
Of course, propagators (and quasi-propagators) corresponding to different particles commute with each other $[\hat {\cal T}_k^{(0)}, \hat {\cal T}_l^{(0)}] = [\hat T_k^{(0)}, \hat T_l^{(0)}] = 0$.

Let us compute the $N$-fermion propagation amplitude on the positive half-line. One finds after some algebra
\begin{eqnarray}\label{Nfermion}
    && {}_A\langle \Vec{x}|\hat {\cal T}_1^{(0)} \hat {\cal T}_2^{(0)} \cdots \hat {\cal T}_N^{(0)}|\Vec{y}\rangle_A =\nonumber\\ 
    && \sum_{\pi\in S_N} (-1)^\pi \prod_{k=1}^N \langle x_k|\hat {\cal T}^{(0)}(\tau)| y_{\pi(k)}\rangle =\nonumber\\ && \det_{ij} \langle x_i|\hat {\cal T}^{(0)}(\tau)| y_j\rangle\nonumber\\{}\nonumber\\
    &&=\left ( \begin{array}{cc}  \det_{ij}G_D(x_i,y_j;\tau) 
     & 0 \\
    0 &  \det_{ij}G_N(x_i,y_j;\tau)
\end{array}\right),
\end{eqnarray}
which tells us that the $N$-particle Dirichlet and Neumann Fermi gases propagate independently on the positive half-line.

For our purpose we need the $N$-fermion quasi-propagator of the mixture of Dirichlet and Neumann particles, for which one obtains the corresponding amplitude as 
\begin{eqnarray}\label{Nfermion-mixture}
    && {}_A\langle \Vec{x}|\hat T_1^{(0)} \hat T_2^{(0)} \cdots \hat T_N^{(0)}|\Vec{y}\rangle_A\nonumber\\&=& \det_{ij} \left(\frac{1}{\sqrt {4\pi\tau}} e^{-\frac{1}{4\tau}(x_i-y_j)^2}\right)\,.
\end{eqnarray}
The diagonal element of the $N$-particle density matrix of these mixed Diriclet-Neumann fermions is therefore
\begin{eqnarray}\label{Nfermion-diagonal}
    {}_A\langle \Vec{x}|\hat T_1^{(0)} \hat T_2^{(0)} \cdots \hat T_N^{(0)}|\Vec{x}\rangle_A = \det_{ij} \left(\frac{1}{\sqrt {4\pi\tau}} e^{-\frac{1}{4\tau}(x_i-x_j)^2}\right)\,,\nonumber\\{}
\end{eqnarray}
which coincides (up to normalization) with \eqref{jpdx} when $V(x)=0$ and upon setting
\begin{equation}\label{Ng}
Ng=\frac{1}{4\tau}\,. 
\end{equation}
This pattern of the $N$-fermion density matrix \eqref{Nfermion-diagonal} on the positive half-line coinciding in form with that of a Fermi gas occupying the whole line with parity-invariant single-particle hamiltonian (the free particle hamiltonian $\hat p^2$ in the present case) will persist also after turning on an external potential $V(x)$ in \eqref{jpdx}. 

The bottom line is that up to a normalization factor, we could have obtained \eqref{Nfermion-diagonal} simply by considering the density matrix of the gas of $N$-free fermions on the {\em whole} real line (for which the single particle hamiltonian $\hat p^2$ is parity invariant), taking its diagonal matrix element in position space, and restricting coordinates of all particles to the positive half-line.   

We close this Appendix with a few words about interparticle spacing statistics, in relation with the discussion in Section \hyperref[subsec:statistics-change]{2.1}. 
For small finite values of $Ng$, this repulsion is of the $\beta=2$ Wigner-Dyson type. Indeed, in the limit $g\rightarrow 0^+$, the integral \eqref{eq:IZ} tends to a constant. Consequently, in this limit, 
\begin{eqnarray}\label{smallg}
\det_{ij}\left(\frac{1}{\sqrt {4\pi\tau}} e^{-\frac{1}{4\tau}(x_i-x_j)^2}\right)&=& \det_{ij} \left(\sqrt {\frac{Ng}{\pi}}\,e^{-Ng(x_i-x_j)^2}\right)\nonumber {}\\ &\propto& g^{\frac{N^2}{2}}\Delta^2(\Lambda^2)\,e^{-2Ng\sum_i x_i^2}\nonumber {}\\
\end{eqnarray}
(up to a $g$-independent normalization factor), which vindicates the assertion. In the opposite limit, as $1/(4\tau)= Ng\rightarrow\infty$ repulsion among fermions disappears completely (over length scales larger than $1/\sqrt{Ng}$). It is natural to expect Poisson level statistics to be associated with this suppressed repulsion. The argument is similar to the one made in Section \hyperref[subsec:statistics-change]{2.1}, and we repeat it here for convenience: Typical consecutive interparticle spacings are of the order $\frac{1}{N}$  or larger. Thus, for large $g\gg N$, the gaussian part of the matrix elements inside the determinant in \eqref{Nfermion-diagonal} becomes essentially the unit matrix $\delta_{ij}$, except for the extremely rare occasion when a finite fraction of all particles condense into much shorter separations of the order of $1/\sqrt{Ng}$. Thus, for $g\gg N$, a good approximation for the joint probability distribution of $N$ fermions associated with \eqref{Nfermion-diagonal} is 
\begin{equation}\label{jpdxiidB}
P(x_1,x_2,\ldots,x_N)\approx\frac{1}{L^N} = {\rm const.}\,,
\end{equation} 
where $L$ is the circumference of the large ring on which the free fermions live. \eqref{jpdxiidB} is further subjected to the constraint that pairs of particles cannot come closer than a distance $\sim 1/\sqrt{Ng}$ in space. That is, except for an excluded volume fraction of the order $N\frac{1}{\sqrt{Ng}} = \sqrt{\frac{N}{g}}\ll 1$, the gas is approximately that of identically independently distributed (i.i.d.) particles, each of which is flatly distributed along the ring. As discussed in Section 
\hyperref[subsec:statistics-change]{2.1}, the unfolded interparticle spacing statistics in the large-$N$ limit is Poissonian $P(s)=e^{-s}$.

As we show in Section \hyperref[subsec:statistics-change]{2.1} many of the features discussed here apply also in the presence of a confining potential $V(x)$ in \eqref{jpdx}.

\section{Gas with \texorpdfstring{$N=2$}{N=2} Particles}\label{AppC}
It is constructive to derive explicit expressions for the particle separation probability $P(s)$ in the gas consisting of two particles, in analogy with Wigner's surmise.  

\begin{figure}[h]
    \centering
    \includegraphics[width=1\linewidth]{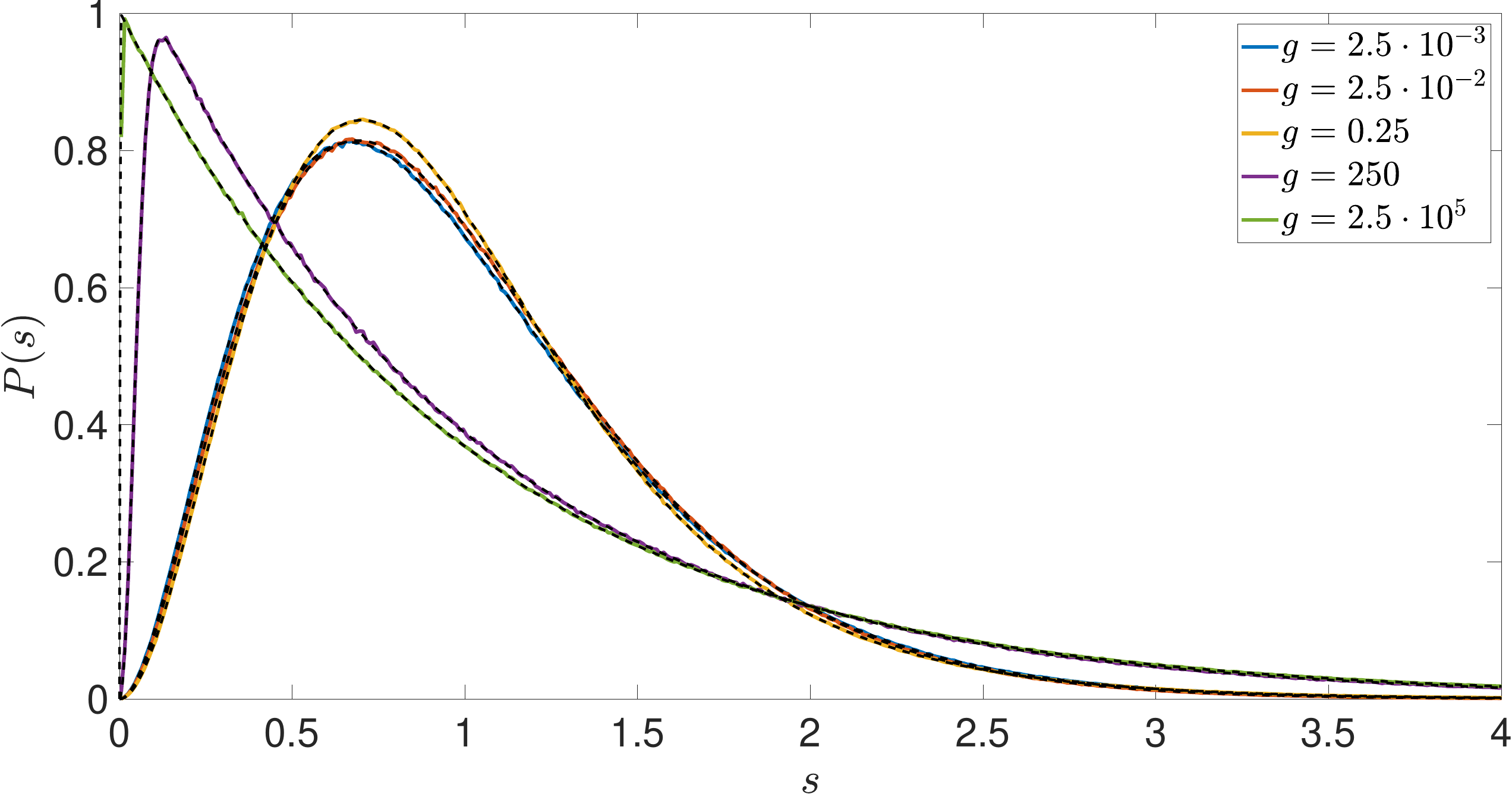}
    \caption{Unfolded spacing distribution of square singular values for Ginibre's potential $V(x)=x$ with $N=2$ for various values of $g$, showing transition from Wigner-Dyson ($g=0$) towards Poisson statistics. Solid lines show numerical data obtained from Monte Carlo simulation of singular values while the overlaying dashes lines demonstrate agreement with the theoretical prediction given in \eqref{PGin}.}
    \label{fig:SVspdistN2}
\end{figure}

For $N=2$ we can write \eqref{jpdx} simply as 
\begin{equation}\label{N=2}
P(x_1,x_2)  = \frac{1}{\mathcal{Z}_2}e^{-2(V(x_1)+ V(x_2))}\left(1-e^{-4g(x_1-x_2)^2}\right)\,.
\end{equation}
Thus, the probability density distribution $P(s) = {\rm Prob}(|x_1-x_2|=s)$ for interparticle separation $s>0$ is 
\begin{eqnarray}\label{Ps2}
P(s) &=& \frac{1}{\mathcal{Z}_2}\left(\int\limits_0^\infty dx_1 P(x_1,x_1+s) + \int\limits_0^\infty dx_2 P(x_2+s,x_2)\right)\nonumber\\{}\nonumber\\
&=&\frac{2}{\mathcal{Z}_2}(1-e^{-4gs^2})
\int\limits_0^\infty dx e^{-2(V(x)+V(x+s))}\,.
\end{eqnarray}
Let us define the single particle probability distribution 
\begin{equation}\label{single}
p(x) = \frac{1}{\cal Z_{\rm sp}} e^{-2V(x)}\,.
\end{equation}
Then, the spacing distribution function for two independent particles identically distributed according to $p(x)$ is 
\begin{equation}\label{Ps2iid}
P_{\rm i.i.d.}(s) = \frac{2}{{\cal Z}^2_{\rm sp}}\int\limits_0^\infty dx e^{-2(V(x)+V(x+s))}\,.
\end{equation}
Therefore, we can rewrite \eqref{Ps2} as 
\begin{equation}\label{Ps22}
P(s) = \frac{{\cal Z}^2_{\rm sp}}{\mathcal{Z}_2}(1-e^{-4gs^2})P_{\rm i.i.d.}(s)
\end{equation}
in which dependence on the potential appears only in the last factor $P_{\rm i.i.d.}(s)$. The $g$ dependence is factored out and sets a length scale $s_0 = \frac{1}{\sqrt{4g}}$ (analogous to the separation scale $s_{\rm micro}$ in Section \hyperref[subsec:statistics-change]{2.1}) for interparticle separations, rendering $P(s)$ effectively independent of $g$ in the regime $s\gg s_0$. In particular, for large $g$, the two particles behave as statistically independent 
\begin{equation}\label{Ps23}
P(s) \approx P_{\rm i.i.d.}(s)
\end{equation}
for $s>s_0\sim~0^+$ (where we have used the fact that ${\mathcal{Z}_2}\approx {\cal Z}^2_{\rm sp}$ in the large $g$ limit). 

For small $g$, on the other hand, $(1-e^{-4gs^2})\approx 4gs^2$, leading to level repulsion similar to the familiar GUE behavior.

For Ginibre's case $V(x)=x$ we have ${\cal Z}_{\rm sp} = \frac{1}{2}$ and $P_{\rm i.i.d.}(s) = 2e^{-2s}$. Thus
\begin{equation}\label{PGin}
P(s)_{\rm Ginibre} = \frac{1}{2{\mathcal{Z}_2}}e^{-2s}(1-e^{-4gs^2})\,.
\end{equation}
For small $g$ it tends to the GUE-like behavior $P(s)_{\rm Ginibre}\propto gs^2 e^{-2s}$. In contrast, for large $g$ and for $s>s_0\sim 0^+$, \eqref{PGin} displays Poisson behavior $P(s)_{\rm Ginibre}\propto e^{-2s}$. These features are displayed (after proper normalization) in Fig.~\ref{fig:SVspdistN2}. We remark that the Poisson behavior occurring already for $N=2$ particles is a special feature of Ginibre's potential. The universal large-$g$ Poisson behavior discussed in Section \hyperref[subsec:statistics-change]{2.1} (following \eqref{jpdxiid}) occurring for arbitrary potential $V(x)$ is in general a large-$N$ effect. 

We can also obtain a simple explicit expression for $V(x)=x^2$. In this case ${\cal Z}_{\rm sp} = \sqrt{\frac{\pi}{8}}$ and $P_{\rm i.i.d.}(s) = \frac{4}{\sqrt{\pi}}e^{-s^2}{\rm erfc}(s)$. Consequently \eqref{Ps2} leads to 
\begin{equation}\label{Pquad}
P(s)_{\rm quadratic} = \frac{\sqrt{\pi}}{2{\mathcal{Z}_2}}(1-e^{-4gs^2})e^{-s^2}{\rm erfc}(s)\,.
\end{equation}
As in the previous case, for small $g$ it exhibits the expected GUE behavior $P(s)_{\rm quadratic}\propto gs^2 e^{-s^2}$.

\section*{Acknowledgements}
\begin{acknowledgments}


J.F.~thanks Larry Schulman for a useful correspondence and Boris Shapiro for a useful discussion. R.R. thanks Micha Riser for manual SIMD vectorization via compiler intrinsics to optimize both data- and instruction-level parallelism in the Monte Carlo simulation code, significantly improving computational efficiency. J.F.~is supported in part by Grant No. 2022158 from the United States-Israel Binational Science Foundation (BSF), Jerusalem, Israel. R.R.~is supported by the Israel Science Foundation through
the Grant No. 956/24.
R.T.S.~acknowledges support from the U.S.~Department of Energy, Office of Science,
Office of Basic Energy Sciences, under Award Number DE-SC0014671.
\end{acknowledgments}

\bibliography{flow}

@article{Challa1986,
  author  = {Challa, Murty S. S. and Landau, D. P. and Binder, K.},
  title   = {Finite-size effects at temperature-driven first-order transitions},
  journal = {Physical Review B},
  volume  = {34},
  pages   = {1841--1852},
  year    = {1986},
  doi     = {10.1103/PhysRevB.34.1841},
  url     = {https://doi.org/10.1103/PhysRevB.34.1841}
}

@article{LeeKosterlitz1991,
  author  = {Lee, Jooyoung and Kosterlitz, J. M.},
  title   = {Finite-size scaling and Monte Carlo simulations of first-order phase transitions},
  journal = {Physical Review B},
  volume  = {43},
  pages   = {3265--3277},
  year    = {1991},
  doi     = {10.1103/PhysRevB.43.3265},
  url     = {https://doi.org/10.1103/PhysRevB.43.3265}
}

@book{NewmanBarkema1999,
  author    = {M. E. J. Newman and G. T. Barkema},
  title     = {Monte Carlo Methods in Statistical Physics},
  publisher = {Oxford University Press},
  address   = {Oxford},
  year      = {1999}
}

@book{LandauBinder2021,
  author    = {David P. Landau and Kurt Binder},
  title     = {A Guide to Monte Carlo Simulations in Statistical Physics},
  edition   = {5},
  publisher = {Cambridge University Press},
  year      = {2021}
}

@article{SwendsenWang1987,
  author  = {R. H. Swendsen and J.-S. Wang},
  title   = {Nonuniversal Critical Dynamics in {M}onte {C}arlo Simulations},
  journal = {Physical Review Letters},
  volume  = {58},
  pages   = {86--88},
  year    = {1987},
  doi     = {10.1103/PhysRevLett.58.86}
}

@article{Evertz1993,
  author  = {H. G. Evertz and G. Lana and M. Marcu},
  title   = {Cluster Algorithm for Vertex Models},
  journal = {Physical Review Letters},
  volume  = {70},
  pages   = {875--879},
  year    = {1993},
  doi     = {10.1103/PhysRevLett.70.875}
}

@article{bhanot1988metropolis,
  title={The {M}etropolis algorithm},
  author={Bhanot, Gyan},
  journal={Reports on Progress in Physics},
  volume={51},
  number={3},
  pages={429--457},
  url={https://iopscience.iop.org/article/10.1088/0034-4885/51/3/003/meta},
  year={1988}
}

@book{krauth2006statistical,
  title={Statistical mechanics: algorithms and computations},
  author={Krauth, Werner},
  volume={13},
  year={2006},
  publisher={OUP Oxford},
  url={https://academic.oup.com/book/54722}
}

@article{abraham1986computational,
  title={Computational statistical mechanics methodology, applications and supercomputing},
  author={Abraham, Farid F},
  journal={Advances in Physics},
  volume={35},
  number={1},
  pages={1--111},
  year={1986},
  publisher={Taylor \& Francis},
  url={https://www.tandfonline.com/doi/abs/10.1080/00018738600101851}
}

@article{chib1995understanding,
  title={Understanding the {M}etropolis-{H}astings algorithm},
  author={Chib, Siddhartha and Greenberg, Edward},
  journal={The {A}merican {S}tatistician},
  volume={49},
  number={4},
  pages={327--335},
  year={1995},
  publisher={Taylor \& Francis},
  url={https://www.tandfonline.com/doi/abs/10.1080/00031305.1995.10476177}
}

@article{scalettar1991ergodicity,
  title = {Ergodicity at large couplings with the determinant {M}onte {C}arlo algorithm},
  author = {Scalettar, Richard T. and Noack, Reinhard M. and Singh, Rajiv R. P.},
  journal = {Phys. Rev. B},
  volume = {44},
  issue = {19},
  pages = {10502--10507},
  numpages = {0},
  year = {1991},
  month = {Nov},
  publisher = {American Physical Society},
  doi = {10.1103/PhysRevB.44.10502},
  url = {https://link.aps.org/doi/10.1103/PhysRevB.44.10502}
}

@article{cohenstead2022fast,
  title = {Fast and scalable quantum {M}onte {C}arlo simulations of electron-phonon models},
  author = {Cohen-Stead, Benjamin and Bradley, Owen and Miles, Cole and Batrouni, George and Scalettar, Richard and Barros, Kipton},
  journal = {Phys. Rev. E},
  volume = {105},
  issue = {6},
  pages = {065302},
  numpages = {22},
  year = {2022},
  month = {Jun},
  publisher = {American Physical Society},
  doi = {10.1103/PhysRevE.105.065302},
  url = {https://link.aps.org/doi/10.1103/PhysRevE.105.065302}
}

@article{feinberg2001single,
  title={`Single ring theorem' and the disk-annulus phase transition},
  author={Feinberg, Joshua and Scalettar, R and Zee, A},
  journal={Journal of Mathematical Physics},
  volume={42},
  number={12},
  pages={5718--5740},
  year={2001},
  publisher={American Institute of Physics},
  url={https://aip.scitation.org/doi/abs/10.1063/1.1412599}
}

@article{feinberg2006Stellenbosch,
  title={Non-Hermitian Random Matrix Theory: summation of planar diagrams, the `single ring' theorem and the disc-annulus phase transition},
  author={Feinberg, Joshua},
  journal={Journal of Physics A: Mathematical and Theoretical},
  volume={39},
  pages={10029-10056},
  year={2006},
  publisher={Institute of Physics},
  url={https://iopscience.iop.org/article/10.1088/0305-4470/39/32/S07}
}

@article{fyodorov2008,
  title={On the mean density of complex eigenvalues for an ensemble of random matrices with prescribed singular values},
  author={Wei, Yi and Fyodorov, Yan V},
  journal={Journal of Physics A: Mathematical and Theoretical},
  volume={41},
  pages={502001},
  year={2008},
  publisher={Institute of Physics},
  url={https://iopscience.iop.org/article/10.1088/1751-8113/41/50/502001}
}

@article{fyodorov2007,
  title={Averages of Spectal Determinants and `Single Ring Theorem' of Feinberg and Zee},
  author={Fyodorov, Yan V and Khoruzhenko, Boris A},
  journal={Acta Physica Polonica B: Mathematical and Theoretical},
  volume={38},
  pages={4067-4077},
  year={2007},
  publisher={Polish Academy of Arts and Sciences},
  url={https://www.actaphys.uj.edu.pl/fulltext?series=Reg&vol=38&page=4067}
}

@article{guionnet2011single,
  title={The single ring theorem},
  author={Guionnet, Alice and Krishnapur, Manjunath and Zeitouni, Ofer},
  journal={Annals of mathematics},
  volume={174},
  pages={1189--1217},
  year={2011},
  publisher={JSTOR},
  url={https://www.jstor.org/stable/pdf/23030522.pdf}
}

@article{Nowak,
  title={Squared eigenvalue condition numbers and eigenvector correlations from the single ring theorem},
  author={Belinschi, Serban and Nowak, Maciej A and Speicher, Roland and Tarnowski, Wojciech},
  journal={Journal of Physics A: Mathematical and Theoretical},
  volume={50},
  pages={105204},
  year={2017},
  publisher={Institute of Physics},
  url={https://iopscience.iop.org/article/10.1088/1751-8121/aa5451}
}

@article{Nowak1,
  title={Complete diagrammatics of the single ring theorem},
  author={Nowak, Maciej A and Tarnowski, Wojciech},
  journal={Physical review E},
  volume={96},
  pages={042149},
  year={2017},
  publisher={American Physical Society},
  url={https://journals.aps.org/pre/abstract/10.1103/PhysRevE.96.042149}
}

@article{feinberg1997non,
  title={Non-{G}aussian non-{H}ermitian random matrix theory: phase transition and addition formalism},
  author={Feinberg, Joshua and Zee, A},
  journal={Nuclear Physics B},
  volume={501},
  number={3},
  pages={643--669},
  year={1997},
  publisher={Elsevier},
  url={https://www.sciencedirect.com/science/article/pii/S0550321397004197}
}

@article{rectangles,
  title={Renormalizing rectangles and other topics in random matrix theory},
  author={Feinberg, J and Zee, A},
  journal={Journal of Statistical Physics},
  volume={87},
  number={May 1997},
  pages={473--504},
  year={1997},
  publisher={Springer},
  url={https://link.springer.com/article/10.1007/BF02181233}
}

@article{Periwal,
  title={Branched polymers from a double-scaling limit of matrix models },
  author={Anderson, A and Myers, R. C. and Periwal, V},
  journal={Nuclear Physics B},
  volume={360},
  number={2-3},
  pages={463--479},
  year={1991},
  publisher={Elsevier},
  url={https://www.sciencedirect.com/science/article/abs/pii/055032139190411P}
}

@article{IZ,
  title={The Planar Approximation. II},
  author={Itzykson, C and Zuber, J.-B.},
  journal={Journal of Mathematical Physics},
  volume={21},
  number={3},
  pages={411--421},
  year={1980},
  publisher={American Institute of Physics},
  url={https://aip.scitation.org/doi/10.1063/1.524438}
}

@article{Parisi,
  title={String Theory on the One Dimensional Lattice},
  author={Parisi, G},
  journal={Physics Letters B},
  volume={238},
  number={2-4},
  pages={213--216},
  year={1990},
  publisher={Elsevier},
  url={https://www.sciencedirect.com/science/article/abs/pii/037026939091723O}
}

@article{MNS,
  title={Generalized Ensembles of Random Matrices},
  author={Moshe, M and Neuberger, H and Shapiro, B},
  journal={Physical Review Letters},
  volume={73},
  number={11},
  pages={1497--1500},
  year={1994},
  publisher={American Physical Society},
  url={https://journals.aps.org/prl/abstract/10.1103/PhysRevLett.73.1497}
}

@article{BIPZ,
  title={Planar Diagrams},
  author={Br\'ezin, E and Itzykson, C and Parisi, G and Zuber, J.-B.},
  journal={Communications in Mathematical Physics},
  volume={59},
  pages={35--51},
  year={1978},
  publisher={Springer-Verlag},
  url={https://link.springer.com/article/10.1007/BF01614153}
}

@article{CMM,
  title={A method of integration over matrix variables: {II}},
  author={Chadha, S and Mahoux, G. and Mehta, M. L.},
  journal={Journal of Physics A: Mathematical and General},
  volume={14},
  pages={579-586},
  year={1981},
  publisher={Institute of Physics},
  url={https://iopscience.iop.org/article/10.1088/0305-4470/14/3/008}
}

@article{BKZ,
  title={Scaling violation in a field theory of closed strings in one physical dimension},
  author={Br\'ezin, E and Kazakov, V. A. and Zamolodchikov, A. l. B.},
  journal={Nuclear Physics B},
  volume={338},
  pages={673-688},
  year={1990},
  publisher={Elsevier},
  url={https://www.sciencedirect.com/science/article/abs/pii/055032139090647V}
}

@book{FH,
    author ="Feynman, R. P. and Hibbs, A. R." ,
    title ="Quantum Mechanics and Path Integrals" ,
    publisher ="McGraw-Hill, Inc." ,
    address = "New York" ,
    year ="1965" ,
}

@book{Schulman,
    author ="Schulman, L. S." ,
    title ="Techniques and Applications of Path Integration" ,
    publisher ="Dover Publications, Inc." ,
    address = "Mineola, NY" ,
    year ="2005" ,
}

@book{Mehta,
  title={Random Matrices (3rd Ed.)},
  author={Mehta, Madan Lal},
  year={2004},
  publisher={Elsevier},
  url={https://www.sciencedirect.com/bookseries/pure-and-applied-mathematics/vol/142/suppl/C}
}

@book{Feller1,
  title={An Introduction to Probability Theory and Its Applications, Vol.1 (3rd Ed.)},
  author={Feller, William},
  year={1968},
  publisher={John Willey \& Sons},
  url={https://www.wiley.com/en-us/An+Introduction+to+Probability+Theory+and+Its+Applications%2C+Volume+1%2C+3rd+Edition-p-9780471257080#download-product-flyer}
}

@book{Feller2,
  title={An Introduction to Probability Theory and Its Applications, Vol.2 (2nd Ed.)},
  author={Feller, William},
  year={1971},
  publisher={John Willey \& Sons},
  url={https://www.wiley.com/en-sg/An+Introduction+to+Probability+Theory+and+Its+Applications%2C+Volume+2%2C+2nd+Edition-p-9780471257097}
}

@book{Vivo,
  title={Introduction to Random Matrices - Theory and Practice},
  author={Livan, G and Novaes, M and Vivo, P.},
  year={2018},
  publisher={Springer Nature},
  url={https://doi.org/10.1007/978-3-319-70885-0}
}

@article{RK-2021,
  title={Power spectrum and form factor in random diagonal matrices and integrable billiards},
  author={Riser, Roman and Kanzieper, Eugene},
  journal={Annals of Physics},
  volume={425},
  pages={168393},
  year={2021},
  publisher={Elsevier},
  url={https://doi.org/10.1016/j.aop.2020.168393}
}

@article{AmeurCC2025,
  title={The two-dimensional {C}oulomb gas: fluctuations through a spectral gap},
  author={Ameur, Yacin and Charlier, Christophe and Cronvall, Joakim},
  journal={Archive for Rational Mechanics and Analysis},
  volume={249},
  number={6},
  pages={63},
  year={2025},
  publisher={Springer},
  url={https://doi.org/10.1007/s00205-025-02133-9} 
}

@article{Chau,
  title={Unitary polynomials in normal matrix models and wave functions for the fractional quantum {H}all effects},
  author={Chau, L.-L. and Yu, Y},
  journal={Physics Letters A},
  volume={167},
  number={5-6},
  pages={452--458},
  year={1992},
  publisher={Elsevier},
  url={https://www.sciencedirect.com/science/article/abs/pii/037596019290604K}
}

@article{HL1,
  title={The Functional Integral on the Half-Line},
  author={Farhi, E. and Gutmann, S},
  journal={International Journal of Modern Physics A},
  volume={5},
  number={15},
  pages={3029--3051},
  year={1990},
  publisher={World Scientific},
  url={https://www.worldscientific.com/doi/10.1142/S0217751X90001422}
}

@article{HL2,
  title={Quantum Mechanics on the Half-Line Using Path Integrals},
  author={Clark, T. E. and Menikoff, R. and Sharp, D. H.},
  journal={Physical Review D},
  volume={22},
  number={12},
  pages={3012--3016},
  year={1980},
  publisher={American Physical Society},
  url={https://journals.aps.org/prd/abstract/10.1103/PhysRevD.22.3012}
}

@article{Vershik,
  title={Asymptotic Theory of Characters of the Symmetric Group},
  author={Vershik, A. M. and Kerov, S.V.},
  journal={Functional Analysis and Its Applications},
  volume={15},
  pages={246--255},
  year={1981},
  publisher={Springer Verlag},
  url={https://doi.org/10.1007/BF01106153}
}

@article{Vershik1,
  title={Asymptotic of the largest and the typical dimensions of irreducible representations of a symmetric group},
  author={Vershik, A. M. and Kerov, S.V.},
  journal={Functional Analysis and Its Applications},
  volume={19},
  pages={21--31},
  year={1985},
  publisher={Springer Verlag},
  url={https://doi.org/10.1007/BF01086021}
}

@article{Logan,
  title={A Variational Problem for Random Young Tableaux},
  author={Logan, B.F. and Shepp, L.A.},
  journal={Advances in Mathematics},
  volume={26},
  number={12},
  pages={206--222},
  year={1977},
  publisher={Elsevier},
  url={https://doi.org/10.1016/0001-8708(77)90030-5}
}

@book{Szabo,
    author ="Szabo, R. J." ,
    title ="Equivariant Cohomology and Localization of Path Integrals, (Section 5.7)" ,
    publisher ="Springer Verlag" ,
    address = "Berlin" ,
    year ="2000" ,
    url = {https://link.springer.com/book/10.1007/3-540-46550-2}
}

@article{Szabo-rev,
  title={Equivariant Localization of Path Integrals, (Section 5.7)},
  author={Szabo, R. J.},
  journal={arXiv [hep-th] 9608068},
  year={1996},
  url={https://arxiv.org/abs/hep-th/9608068}
}

@article{rider2003limit,
  title={A limit theorem at the edge of a non-{H}ermitian random matrix ensemble},
  author={Rider, Brian},
  journal={Journal of Physics A: Mathematical and General},
  volume={36},
  number={12},
  pages={3401--3409},
  year={2003},
  url={https://doi.org/10.1088/0305-4470/36/12/331}
}

@article{bender2010edge,
  title={Edge scaling limits for a family of non-{H}ermitian random matrix ensembles},
  author={Bender, Martin},
  journal={Probability theory and related fields},
  volume={147},
  number={1},
  pages={241--271},
  year={2010},
  publisher={Springer},
  url={https://doi.org/10.1007/s00440-009-0207-9}
}

@article{cipolloni2021edge,
  title={Edge universality for non-{H}ermitian random matrices},
  author={Cipolloni, Giorgio and Erd{\H{o}}s, L{\'a}szl{\'o} and Schr{\"o}der, Dominik},
  journal={Probability Theory and Related Fields},
  volume={179},
  number={1},
  pages={1--28},
  year={2021},
  publisher={Springer},
  url={https://doi.org/10.1007/s00440-020-01003-7}
}

@article{campbell2025spectral,
  title={On the spectral edge of non-{H}ermitian random matrices},
  author={Campbell, Andrew and Cipolloni, Giorgio and Erd{\H{o}}s, L{\'a}szl{\'o} and Ji, Hong Chang},
  journal={The Annals of Probability},
  volume={53},
  number={6},
  pages={2256--2308},
  year={2025},
  publisher={Institute of Mathematical Statistics},
  url={https://doi.org/10.1214/25-AOP1761}
}

@article{akemann2025spectral,
  title={Spectral Density and Eigenvector Nonorthogonality in Complex Symmetric Random Matrices},
  author={Akemann, Gernot and Fyodorov, Yan V and Savin, Dmitry V},
  journal={arXiv preprint arXiv:2511.21643},
  year={2025},
  url={https://arxiv.org/abs/2511.21643}
}

@inproceedings{liu2026repeated,
  title={Repeated Erfc Statistics for Deformed {GinUEs}},
  author={Liu, Dang-Zheng and Zhang, Lu},
  booktitle={Annales Henri Poincar{\'e}},
  volume={27},
  number={4},
  pages={1165--1205},
  year={2026},
  organization={Springer},
  url={https://doi.org/10.1007/s00023-025-01561-3}
}

@article{AmeurHedenmalmMakarov2015,
author = {Yacin Ameur and Haakan Hedenmalm and Nikolai Makarov},
title = {{Random normal matrices and Ward identities}},
volume = {43},
journal = {The Annals of Probability},
number = {3},
publisher = {Institute of Mathematical Statistics},
pages = {1157 -- 1201},
year = {2015},
doi = {10.1214/13-AOP885},
URL = {https://doi.org/10.1214/13-AOP885}
}

@article{LR2016,
  title={Fine asymptotic behavior for eigenvalues of random normal matrices: ellipse case},
  author={Lee, Seung-Yeop and Riser, Roman},
  journal={Journal of Mathematical Physics},
  volume={57},
  number={2},
  year={2016},
  publisher={AIP Publishing},
  url={https://doi.org/10.1063/1.4939973}
}

@article{AKM2019,
  title={Rescaling Ward identities in the random normal matrix model},
  author={Ameur, Yacin and Kang, Nam-Gyu and Makarov, Nikolai},
  journal={Constructive Approximation},
  volume={50},
  number={1},
  pages={63--127},
  year={2019},
  publisher={Springer},
  url={https://doi.org/10.1007/s00365-018-9423-9}
}

@article{HW2022,
  title={Planar orthogonal polynomials and boundary universality in the random normal matrix model},
  author={Hedenmalm, H{\aa}kan and Wennman, Aron},
  journal={Acta Mathematica},
  volume={227},
  number={2},
  pages={309--406},
  year={2022},
  publisher={International Press of Boston},
  url={https://doi.org/10.4310/ACTA.2021.v227.n2.a3}
}

@article{cronvall2025direct,
  title={A direct approach to soft and hard edge universality for random normal matrices},
  author={Cronvall, Joakim and Wennman, Aron},
  journal={arXiv preprint arXiv:2511.18628},
  year={2025},
  url={https://arxiv.org/abs/2511.18628}
}

@article{charlier2025smallest,
  title={Smallest gaps of the two-dimensional {C}oulomb gas},
  author={Charlier, Christophe},
  journal={arXiv preprint arXiv:2507.23502},
  year={2025},
  url={https://arxiv.org/abs/2507.23502}
}

@article{Chalker1996,
  title={Spectral Rigidity and Eigenfunction Correlations at the Anderson Transition},
  author={Chalker, J.T. and Kravtsov, V.E. and Lerner, I.V.},
  journal={JEPT Letters},
  volume={64},
  pages={386-392},
  year={1996},
  publisher={Springer Nature},
  url={https://doi.org/10.1134/1.567208}
}

@article{Kravtsov1,
  title={Energy level statistics of a critical random matrix ensemble},
  author={Ndawana, M.L. and Kravtsov, V.E.},
  journal={Journal of Physics A: Mathematical and General},
  volume={36},
  pages={3632},
  year={2003},
  publisher={Institute of Physics},
  url={https://iopscience.iop.org/article/10.1088/0305-4470/36/12/344}
}

@article{Kravtsov2,
  title={Level compressibility in a critical random matrix ensemble: the second virial coefficients},
  author={Kravtsov, V.E. and Yevtushenko, O. and Cuevas, E},
  journal={Journal of Physics A: Mathematical and General},
  volume={39},
  pages={2021},
  year={2006},
  publisher={Institute of Physics},
  url={https://iopscience.iop.org/article/10.1088/0305-4470/39/9/003},
  note = {Corrigendum: \emph{Journal of Physics A: Mathematical and Theoretical {\bf 44}(2011)189501}}
}

\end{document}